\documentclass[a4paper,fleqn,usenatbib]{mnras}

\usepackage[T1]{fontenc}
\usepackage{ae,aecompl}

\usepackage{natbib}

\usepackage{graphicx}        
\usepackage{amsmath}        
\usepackage{amssymb}        
\usepackage{color}          

\defcitealias{riffelA16}{R16}
\defcitealias{colina00}{C00}
\defcitealias{colina02}{C02}
\defcitealias{maraston05}{M05}
\defcitealias{bc03}{BC03}
\defcitealias{conroy09}{C09}
\defcitealias{maraston11}{M11}

\title[SPs in the nuclear region of NGC\,4303]{A SINFONI view of the nuclear activity and circumnuclear star formation in NGC\,4303 - II: Spatially resolved stellar populations}

\author[N. Z. Dametto et al.]{
Natacha Z. Dametto,$^{1}$\thanks{E-mail: natacha.zanon@ufrgs.br}
R. Riffel,$^{1}$ 
L. Colina,$^{2}$
R. A. Riffel,$^{3}$ 
J. Piqueras L\'opez,$^{2}$ 
\newauthor
R. I. Davies,$^{4}$
L. Burtscher,$^{5}$
R. B. Menezes,$^{6}$
S. Arribas,$^{2}$
M. G. Pastoriza,$^{1}$
A. Labiano,$^{7}$
\newauthor
T. Storchi-Bergmann,$^{1}$
L. G. Dahmer-Hahn,$^{1}$
D. A. Sales$^{8}$
\\
$^{1}$Departamento de Astronomia, Instituto de F\'isica, Universidade Federal do Rio Grande do Sul, CP\,15051, Porto Alegre, RS\,91501-970, Brazil\\
$^{2}$Centro de Astrobiolog\'ia (CAB, CSIC-INTA), Carretera de Ajalvir, E-28850 Torrej\'on de Ardoz, Madrid, Spain\\
$^{3}$Departamento de F\'isica, Centro de Ci\^encias Naturais e Exatas, Universidade Federal de Santa Maria, Santa Maria, RS\,97105\-900, Brazil\\
$^{4}$Max Planck Institut f\"ur extraterrestrische Physik, Postfach 1312, D-85741, Garching, Germany\\
$^{5}$Leiden Observatory, Leiden University, PO Box 9513, NL2300 RA Leiden, the Netherlands\\
$^{6}$Instituto de Astronomia Geof\'isica e Ci\^encias Atmosf\'ericas, Universidade de S\~ao Paulo, Rua do Mat\~ao 1226, S\~ao Paulo, SP Brazil\\
$^{7}$Institute for Astronomy, Department of Physics, ETH Zurich, CH-8093 Zurich, Switzerland\\
$^{8}$Instituto de Matem\'atica, Estat\'istica e F\'isica, Universidade Federal do Rio Grande, Rio Grande 96203-900, Brazil.}
 
\date{Accepted XXX. Received YYY; in original form ZZZ}

\pubyear{2017}

\begin{document}
\label{firstpage}
\pagerange{\pageref{firstpage}--\pageref{lastpage}}
\maketitle

\begin{abstract}
We present a spatially resolved stellar population study of the inner $\sim$200\,pc radius of NGC\,4303 based on near-infrared integral field spectroscopy with SINFONI/VLT at a spatial resolution of 40-80\,pc and using the {\sc starlight} code. We found the distribution of the stellar populations presents a spatial variation, suggesting an age stratification. Three main structures stand out. Two nuclear blobs, one composed by young stars (t $\leq$ 50\,Myr) and one with intermediate-age stars (50\,Myr $<$ t $\leq$ 2\,Gyr) both shifted from the centre. The third one is an internal intermediate-age spiral arm-like structure, surrounding the blob of young stars.
Our results indicate star formation has occurred through multiple bursts in this source. Furthermore, the youngest stellar populations (t $\lesssim$ 2\,Gyr) are distributed along a circumnuclear star-forming ring with r$\sim$250\,pc. The ring displays star formation rates (SFRs) in the range of 0.002-0.14\,M$_{\odot}$yr$^{-1}$, favoring the `pearls-on-a-string' scenario. The old underlying bulge stellar population component (t $>$ 2\,Gyr) is distributed outside the two blob structures. For the nuclear region (inner $\sim$60\,pc radius) we derived a SFR of 0.43\,M$_{\odot}$yr$^{-1}$ and found no signatures of non-thermal featureless continuum and hot dust emission, supporting the scenario in which a LLAGN/LINER-like source is hidden in the centre of NGC\,4303. Thus, our results reveal a rather complex star formation history in NGC\,4303, with different stellar population components coexisting with a low efficiency accreting black hole in its centre.
\end{abstract}




%

\section{Introduction}\label{intro}


The stellar population synthesis technique is a powerful tool to derive the galaxy star formation history (SFH). Disentangling the intrinsic properties of the galaxy, such as mass, age, metallicity and dust is a key step for the understanding of a galaxy formation and evolution.  For example, there are physical properties of the galaxy bulges that correlate with properties of the active galactic nuclei (AGN) they harbor, e.g. the M$_{\bullet} - \sigma_{\star}$ relation \citep{ferrarese00,gebhardt00a,kormendy13}, in which the mass of the supermassive black hole (SMBH, M$_{\bullet}$) correlates with the velocity dispersion of the bulge stars ($\sigma_{\star}$). The existence of such correlations suggests a possible causal link between the bulge formation and the central black hole or even that the evolution of both might be regulated by a common effect. A strong candidate to explain the relation between these two phenomena is circumnuclear star formation, since they both depend on the inflow of gas towards the innermost regions of the galaxy \citep[e.g.][]{shlosman90,combes94,rosario2018}.

Star formation tracers in the optical spectral range, for example, are considerably well known and have been used to identify star formation in galaxies over the years \citep[e.g., ][]{kennicutt88,bica88,worthey97,gu06}. Over the past two decades, optical studies on scales of hundreds of parsecs around the nucleus of Seyfert galaxies have shown that in $\sim$40\% of them, young stars (t $\lesssim$ 50\,Myr) coexist with the AGN \citep[e.g.][]{storchi00,storchi01,gonzalez01,cid04,asari07,dors08}, supporting the so-called AGN-starburst connection \citep{shlosman89,shlosman90,heckman97}. Moreover, these studies suggest the main difference between the stellar population of active and non-active galaxies is an excess of intermediate-age stars (t $\sim$0.05$-$2\,Gyr) in the former. 

Similar results were found in stellar population studies using near-infrared (NIR) long-slit spectroscopy \citep{riffel07,riffel09,martins10}, suggesting the continuum of Seyfert are dominated by the contribution of intermediate-age stellar populations. Moreover, contribution of hot dust emission to the nuclear NIR spectra was found in 50\% of Seyfert\,1 and in 20\% of Seyfert\,2 sources \citep{rodriguezA05a,rodriguezA06,riffelA09a,riffelA09b,riffel09}. The fact that the NIR is less affected by dust than the optical bands makes this spectral range the most suitable one to unveil the nuclear stellar populations in highly obscured sources \citep{origlia00}, such as the case of Seyfert\,2 galaxies. The observed properties of this subclass of AGN suggest the SMBH powering the AGN is obscured to the line of sight by the torus surrounding the central engine. In addition, as has already been shown by e.g. \citet{davies06}, star formation can take place within the torus, on scales of tens of parsecs from the nucleus. 

The advent of {\it James Webb Space Telescope} (JWST) and the near-and mid-infrared integral field spectroscopic capabilities it will provide, opens the study of spatially resolved obscured stellar populations and hot dust using the entire 0.6 to 28$\mu$m range. Thus, to test the stellar population synthesis method in this particular spectral range is of fundamental importance for forthcoming studies with JWST.

The AGNIFS (AGN Integral Field Spectroscopy) team has started to characterize the stellar population in the inner kiloparsecs of a sample of nearby Seyfert galaxies (Mrk\,1066, Mrk\,1157, NGC\,1068, NGC\,5548 and Mrk\,573) using {\it Gemini Near-Infrared Integral Field Spectrograph} (NIFS). For Mrk\,1066, Mrk\,1157 and Mrk\,573 the inner $\sim$200\,pc are dominated by old stars ($t \geq 2$\,Gyr), while intermediate-age stars ($0.3 \leq t \leq 0.7$\,Gyr) are the dominant contributors to  the circumnuclear rings found in these sources \citep{riffelA10,riffel11c,diniz17}. These circumnuclear rings of intermediate-age stars are correlated with low stellar velocity dispersion values ($\sigma_{\star} \sim 50$\,kms$^{-1}$), being consistent with a scenario in which the origin of the low-$\sigma_{\star}$ rings is a past event which triggered an inflow of gas and formed stars which still keep the colder kinematics of the gas from which they formed. \citet{storchi12} found two recent episodes of star formation in NGC\,1068: a first one that took place 300\,Myr ago extending over the inner 300\,pc of the galaxy and a second one that occurred just 30\,Myr ago in a ring-like structure at $\sim$100\,pc from the nucleus, where it is coincident with an expanding ring of warm H$_2$ emission. \citet{schonell17}, on the other hand, detected a dominant intermediate-age stellar population component (SPC) in the inner 160\,pc of NGC\,5548, while an old ($> 2$\,Gyr) SPC dominates the region between 160 and 300\,pc. Dust emission has been detected in all sources (with exception of NGC\,1068) accounting for 30 to 90\% of the K-band nuclear flux, while a featureless continuum component associated with the AGN emission was detected in three sources, contributing with $\sim$20$-$60\% of the K-band nuclear flux (NGC\,1068, Mrk\,573 and NGC\,5548).

Modeling the Br$\gamma$ equivalent width (EW), supernovae rate and mass-to-light ratio, \citet{davies07} have quantified the SFH in the centre of nine nearby Seyfert galaxies using their {\sc stars} code. Their results indicate the age of the stars which contributes most to the NIR continuum lie in the range 10-30\,Myr, pointing out these ages should be considered only as `characteristic', as they have not performed a proper spectral synthesis, suggesting there may be simultaneously two or more stellar populations that are not coeval.

All these previous studies have focused on either nearby luminous Seyfert galaxies, in which the output energy is dominated by the AGN, or on galaxies with luminous circumnuclear star-forming rings. Low-luminosity AGNs (LLAGNs), on the other hand, including low-luminosity Seyfert galaxies, classical LINERs, weak-[O{\sc i}] LINERs and LINER/H{\sc ii} transition-like objects are the most common type of galaxies that display nuclear activity \citep[for a review on LINERs, see][]{filippenko03,singh13,belfiore16,hsieh17}. LINERs alone comprise 50\% to 70\% of AGNs and 20\% to 30\% of all galaxies in surveys of nearby bright galaxies \citep{ho97}. Therefore, to identify the nature of the energy source in LLAGNs, as well as to disentangle the contribution of both SPCs and accreting black holes to the energy output of these sources is of utmost importance.
In this context, a key object to move forward on the understanding of the role played by star formation in LLAGNs is NGC\,4303, a nearby LINER/Seyfert\,2 \citep{filippenko86,kennicutt89,ho97,colina99} galaxy with a LLAGN coexisting with a young massive star cluster in the nucleus and a star-forming circumnuclear ring (see Sec.~\ref{previous}). 

\subsection{Previous stellar population studies of NGC4303}\label{previous}

NGC\,4303 is a barred spiral classified as SB(rs)bc \citep{vaucouleurs91}, located in the Virgo Cluster, at a distance of 16.1\,Mpc \citep{ferrarese96}. A nuclear UV-bright spiral arm with an outer radius of 225\,pc and spiraling all the way down to the unresolved (size $<$ 8\,pc) UV-bright core was detected in the HST/WFPC2 F218W images \citep{colina97}. This spiral structure was traced by several distinct regions, identified as young massive stellar clusters. These authors concluded the unresolved LINER-like core contributes with 16\% of the UV luminosity of NGC\,4303, which is dominated by the massive star-forming regions. Another important result is the presence of spiral star-forming structure in the nuclear region of a barred spiral supports the bar-induced AGN-starburst scenario. In this sense, the gas in the bar-driven nuclear spiral can flow inward towards the core and fuel a preexisting black hole, producing an AGN \citep{shlosman90,combes03}.

In a later work, \citet{colina99} used optical integral field spectroscopy to study the inner 9" $\times$ 8" (i.e. 700$\times$620\,pc$^2$) and found, from optical emission-line ratios, that the circumnuclear star-forming regions (CNSFRs) of this source have ages of 2$-$3\,Myr (extremely young clusters of massive stars). Also, they concluded the presence of a massive (8 $\times$ 10$^4$M$_{\odot}$) and young (3.5$-$4\,Myr) stellar cluster in the nucleus is consistent with the observed properties of the core of NGC\,4303 (e.g. optical emission-line ratios, UV and H$\alpha$ luminosities). Nevertheless, the presence of a non-thermal power-law AGN-like ionizing source cannot be ruled out. Thus, the authors classified the core of this galaxy as a [O{\sc i}]-weak LINER or low-excitation Seyfert\,2.

Studying the radial distributions of the Mg$_2$ and Fe5270 Lick spectral indices in the disk of NGC\,4303, \citet{molla99} found while in NGC\,4303 both indices steeply rise towards the central region ($\lesssim$~2\,kpc), in the other objects studied by them (NGC\,4321 and NGC\,4535) a central dip is observed. In their study, by using SSP models, these authors inferred that NGC\,4303 is still forming stars. 

Using WFPC2 (F218W,F606W) and NICMOS (F160W) {\it Hubble Space Telescope} (HST) images, \citet[][hereafter C00]{colina00} studied the inner 300\,pc of NGC\,4303 and identified a nuclear elongated bar-like structure of 250\,pc in size. The images revealed a complex gas/dust distribution with a two-arm spiral structure of about 225\,pc in radius. Also, they estimated the age of the UV-bright knots located along the star-forming spiral structure as 5$-$25\,Myr and masses of 0.5$-$1 $\times$ 10$^5$M$_{\odot}$. 

From UV imaging and spectroscopy (HST/STIS), \citet[][hereafter C02]{colina02} concluded the UV emission from the nucleus of NGC\,4303 comes from a region of 3.1\,pc and it is identified as a young (4\,Myr), massive (10$^5$\,M$_{\odot}$) nuclear cluster. The authors found this compact super-star cluster (SSC), commonly detected in the (circum)nuclear regions of spirals and starburst galaxies, is the dominant ionizing source in the nucleus. According to this study, an additional non-thermal ionizing source due to an AGN is not required. They also discuss the possibility of having an intermediate/old (1$-$5\,Gyr) star cluster coexisting with the low efficiency accreting black hole and the young and luminous SSC. 
\citet{bailon03} analyzed 2$-$10\,keV observations from the Chandra X-ray satellite and indicated an additional compact source was required to explain the 1.5$-$5\,keV emission, possibly a low-luminosity AGN. 

In \citet[][hereafter R16]{riffelA16} we presented the results concerning the kinematics and excitation properties of the different phases of the interstellar medium in the circumnuclear region of NGC\,4303 using the same SINFONI datacubes presented in this work. A circumnuclear ring of star-forming regions (r$\sim$200$-$250\,pc) was detected, displaying young ages in the range 2.5$-$15\,Myr. Star formation in the ring appears to be episodic, with stars forming quasi-simultaneously. Moreover, NIR emission-line ratios ([Fe{\sc ii}]/Br$\gamma$ and H$_2$/Br$\gamma$) are consistent with the presence of an AGN and/or a SN-dominated star-forming region in the core (inner 60\,pc radius) of NGC\,4303. 


Here we perform, for the first time, a spatially resolved full spectral fitting stellar population study of the nuclear region ($\sim$200\,pc radius) of NGC\,4303, using both {\sc starlight} code and VLT/SINFONI data. This paper is structured as follows: Sec.~\ref{data} presents the observations and data reduction procedures, while in Sec.~\ref{method} we introduce the stellar population synthesis method used in this work. The results are presented in Sec.~\ref{results} and discussed in Sec.~\ref{discussion}. Finally we present our conclusions in Sec.~\ref{conclusions}. North is up and east is to the left throughout the images and maps presented in this paper.


\section{Observations and Data Reduction}\label{data}

The data used are the same as the datacubes presented in \citetalias{riffelA16}. Here we summarize the information on observations and data reduction process, as follows. The observations were done during the period 82B (February 2009) at the ESO {\it Very Large Telescope} (VLT) with SINFONI, a NIR integral field spectrograph \citep{eisenhauer03,bonnet04}. The pointings were centered on the nucleus of the galaxy, covering a field of view (FoV) of $\sim$8\arcsec $\times $8\arcsec per exposure, enlarged by dithering up to $\sim$9\farcs25 $\times$ 9\farcs25, with a plate scale of 0\farcs125$\times$0\farcs250 pixel$^{-1}$. The final data cube was re-sampled to a scale of 0\farcs125$\times$0\farcs125 pixel$^{-1}$, corresponding to a spatial sampling of $\sim$10\,pc per spaxel.

The data were taken in the J (1.10$-$1.35 $\mu$m), H (1.45$-$1.80 $\mu$m), and K (1.97$-$2.44 $\mu$m) bands with a total integration time of 2400\,s per band. In the same way, a set of photometric standard stars was observed to perform the telluric and flux calibration. We estimated the spatial resolution of our seeing-limited observations by fitting a 2D Gaussian profile to a collapsed image of the standard stars. The spatial resolution (FWHM) measured for each band is $\sim$1\arcsec, $\sim$0\farcs6, and $\sim$0\farcs5 for J-, H-, and K-band, respectively, that correspond to 78\,pc, 47\,pc, and 39\,pc at the adopted distance of 16.1\,Mpc for NGC\,4303. The reduction and calibration processes were performed using the standard ESO pipeline {\sc esorex} (version 3.8.3), and our own IDL routines \citepalias[see][where a complete description of the reduction and calibration procedures is provided]{riffelA16}.


 \subsection{Relative flux calibration}\label{irtf}

The accurate determination of the continuum shape plays an important role on the stellar population determinations \citep{baldwin18}. The stellar population synthesis technique used in this work is strongly dependent on a reliable continuum fit, meaning that any problem in the relative flux calibration between spectral bands would compromise the results. When dealing with emission-line measurements, on the other hand, this issue is much less important. The continuum estimation needed to fit an emission line is done within the range of only one spectral band, being nearly independent of relative flux calibrations. With this in mind, we decided to add an extra step concerning the relative flux calibration of our data set, in order to get a reliable relative calibration, adequate for the purpose of spectral energy distribution fitting.

Our target was observed in three different nights and only a limited number of standard stars were available to perform the flux calibration. Although the efficiency curves appear to be consistent, the flux calibration of the J-band data might be inaccurate outside the 1.15$-$1.30$\mu$m range, since the slope of the curves is very uncertain beyond this limit and the spectra is completely dominated by the noise from the atmospheric absorption bands. This fact is translated to uncertainties on the slope of the spectra in the data cubes (DCs) which, thus, cannot be used to perform a reliable stellar population synthesis using full spectral fitting \citep[see][]{cid05a,baldwin18}.
 
To address that, we used NIR cross-dispersed data of NGC\,4303 from \citet{martins13a}, obtained at the NASA 3m Infrared Telescope Facility (IRTF) using the SpeX spectrograph. As the observations at the J-, H- and K-band are done simultaneously in the cross-dispersed mode, the data do not suffer from relative flux calibration problems. Therefore, assuming the shape of the SpeX data was correct, we used it to scale our SINFONI observations. It is worth mentioning we have used SINFONI integrated spectra extracted with the same aperture size and position angle as the SpeX data.


%

\subsection{Instrumental fingerprint removal}\label{fp}

The instrumental fingerprint removal was performed using the Principal Component Analysis (PCA) Tomography technique, which consists of applying PCA to data cubes. First, the spectral lines of the data cube were removed. Then, PCA was applied and the obtained eigenvectors related to the fingerprint were used to construct a data cube containing only the fingerprint. Such a data cube was subtracted from the original one, completing the removal of the instrumental artifact. This entire procedure was applied separately to the J, H and K bands of the data cube. For more details, see \citep{menezes14,menezes15}.

\section{stellar population synthesis method}\label{method}

A common way to disentangle the spectral energy distribution components of a galaxy spectrum is by performing stellar population synthesis. This method consists in fitting the galaxy absorption and continuum spectrum with a combination of simple stellar population (SSP) components. Therefore, the two main ingredients are: i) the SSP templates (hereafter, base set) and ii) the fitting code. 

\subsection{Base Set}

An ideal set of templates should be able to foresee all the features expected to be found in spectra of galaxies \citep{schmidt91,cid05b}. In other words, a reliable base set would be an empirical library of integrated spectra of star clusters \citep[i.e. they only depend on ages and metallicities of the stars and are free from any assumptions on stellar evolution and the initial mass function - ][]{bica86,riffel11a}. However, up to now there is no such library available in the literature for the NIR spectral region. Thus, the use of a base set composed of theoretical SSPs, covering this spectral region, has become a common approach \citep[e.g.][]{riffel09,martins10,dametto14}. 

Since the NIR carries fingerprints from evolved stars \citep[e.g.][]{riffel07,ramos09,martins13b, riffel15} and these are crucial to model the absorption line spectra of the galaxies, it is important to make use of SSPs models that can predict these features. Following \citet{dametto14}, we decided to used the \citet[][hereafter M05]{maraston05}\footnote{Available at http://www.icg.port.ac.uk/$\sim$maraston/Claudia's\_ Stellar\_Population\_Model.html} Evolutionary Population Synthesis (EPS) models, which include empirical spectra of C- and O-rich stars \citep{lw00} and thus, are able to predict these features. We also tested other SSPs models from \citet[][hereafter BC03]{bc03} and \citet[][hereafter C09]{conroy09}, see Sec.~\ref{mc} for this discussion.

The \citetalias{maraston05} models span an age range from 0.001\,Myr to 15\,Gyr according to a grid of 67 models with six different metallicities (0.005 $\leq$ Z/Z$\sun$ $\leq$ 3.5), 2 Initial Mass Functions (IMFs) (Salpeter and Kroupa) and 3 horizontal branch (HB) morphologies \citepalias[red, intermediate or blue - for more details, see][]{maraston05}. It is worth noting the models with Z=0.005Z$\sun$ and Z=3.5Z$\sun$ are provided only for ages older than 1\,Gyr (in a grid of 16 ages) and are based on Cassisi \citep{cassisi97a,cassisi97b} and Padova 2000 \citep{girardi00} isochrones, respectively. The remaining 4 metallicities, (computed for the full grid of 67 ages) are associated with Cassisi + Geneva \citep{schaller92} tracks. The stellar spectra were taken from the BaSeL 2.2 library \citep{lejeune97,lejeune98}, covering the spectral range of 91$\textrm{\AA}$ to 160$\mu$m, with a spectral resolution of 5$-$10\,$\textrm{\AA}$ up to the optical region, and 20$-$100\,$\textrm{\AA}$ in the near- and far-infrared.

The BaSeL 2.2 is a library of low-resolution stellar spectra based on the theoretical templates compiled by \citet{lejeune97,lejeune98}. This library is widely used in stellar population synthesis studies and was constructed by combining the model atmosphere spectra of \citet{bessell89,bessell91} with the models for cool stars by \citet{fluks94}. As synthetic spectral libraries do not contain TP-AGB carbon-and oxygen-rich stars, empirical and time-averaged spectra of C- and O-type stars from \citet{lancon02} were included in the \citetalias{maraston05} models.

In order to avoid redundant information and degeneracies in the base set, we used only the representative SSPs \citep[see][for further details]{dametto14} and ended up with a final base set composed as follows: 31 ages (1.0\,Myr $\leq$ t $\leq$ 13.0\,Gyr) for each of the 4 metallicities (Z= 0.02, 0.5, 1 and 2 $Z_{\odot}$) totaling 124 SSPs. We also included blackbody functions for temperatures in the range 700$-$1400\,K in steps of 100\,K and a power law ($F_{\nu} \propto \nu^{-1.5}$) in order to account for possible contributions from dust emission and from a featureless continuum, respectively, at the nucleus \citep{cid04,riffel09}.


The spectral resolution of \citetalias{maraston05} models in the NIR is significantly lower (R $\leq$ 250) than that of the observed data (R$\sim$ 2000) and varies with wavelength. For this reason, observations were degraded to the models' resolution by convolving them with a Gaussian. 


\subsection{Fitting Code}\label{code}

Other fundamental ingredient in stellar population fitting is the code. Following \citet{dametto14}, we used {\sc starlight} code \citep[][]{cid05a,mateus06}, which fits the observed spectrum {\it $O_{\lambda}$} with a combination in different proportions of {\it $N_{\star}$} SSPs in the base set -- {\it $b_{j, \lambda}$} -- taken from the EPS models. One of the key features of {\sc starlight} is that the code fits the entire spectrum (from 0.8 to 2.4 $\mu$m in this case), excluding emission lines and spurious features (e.g. cosmic rays, and telluric regions), which are masked out. 
 
Basically, {\sc starlight} solves the following equation for a model spectrum {\it $M_{\lambda}$} \citep{cid05b}:

\begin{equation}
{\it M_{\lambda}= M_{\lambda 0} \left[ \sum_{j=1}^{N_{\star}} x_j b_{j, \lambda} r_{\lambda} \right] \otimes G(v_{\star}, {\sigma}_{\star})},
\end{equation}

\noindent where ${\it M_{\lambda 0}}$ is the synthetic flux at the normalization wavelength ($\lambda_0$=2.067$\mu$m); ${\it x_j}$ is the {\it j}th population vector component of the base set; ${\it b_{j, \lambda} r_{\lambda}}$ is the reddened spectrum of the {\it j}th SSP normalized at ${\lambda}_0$ in which ${\it r_{\lambda}=10^{-0.4(A_{\lambda}-A_{\lambda 0})}}$ is the extinction term; ${\it \otimes}$ denotes the convolution operator and ${\it G(v_{\star}, {\sigma}_{\star})}$ is the gaussian distribution used to model the line-of-sight stellar motions, centered at velocity ${\it v_{\star}}$ with dispersion ${\it {\sigma}_{\star}}$. We choose as normalization wavelength $\lambda_0$=2.067$\mu$m, since K-band spectra present a higher SNR than those in the J- and H-band, and the spectral region near 2.067$\mu$m is free from emission/absorption lines. 

Velocity dispersion is a free parameter for {\sc starlight} which broadens the SSPs in order to better fit the absorption lines in the observed spectra, however this step is not relevant in our case. Assuming \citetalias{maraston05} (BaSeL based models) spectral resolution in velocity units as $\sim$1500\,kms$^{-1}$, the velocity dispersion is $\sim$640\,kms$^{-1}$, which are both much higher values than those calculated in \citetalias{riffelA16} ($v_{\star}$:-80/+80\,kms$^{-1}$ and $\sigma_{\star}$: 20/100\,kms$^{-1}$) for the stellar kinematics. Thus, because of the models' low resolution, we kept the kinematic parameters fixed ($v_{\star}$=0.0 and $\sigma_{\star}$=640\,kms$^{-1}$) during the fits. It is important to highlight that for the low resolution models, the age information is encoded in the continuum shape \citep{riffel09,baldwin18,dahmer18}.

The extinction law used in this work was that of Calzetti law \citep{calzetti00} implemented by Hyperz \citep{bol00}, a public photometric redshift code which computed the Calzetti extinction law for $\lambda >$  2.2$\mu$m. 

Lastly, the code searches for the minimum of the equation:

\begin{equation}
\label{chi2}
{\it {\chi^2 = \sum_{\lambda}^{} {\left[ \left( {\it O_{\lambda}-M_{\lambda}}\right)w_{\lambda} \right]}^2 }},
\end{equation}

\noindent and the best fit is achieved. In order to measure the robustness of the stellar population fit, we can use {\sc starlight} output parameters ${\chi}^2$ and Adev. The later is the percent mean deviation $|O_{\lambda} - M_{\lambda}|/O_{\lambda}$, where $O_{\lambda}$ is the observed spectrum and $M_{\lambda}$ is the fitted model. Emission lines and spurious features (telluric regions, cosmic rays) are masked out by using ${\it w_{\lambda}=0}$ in the regions where they are located. This procedure is done by first constructing a general mask, based on the emission-lines position. Next, we inspect spaxel-by-spaxel to remove any additional spurious data. For more details see \citet{dametto14} and {\sc starlight} manual available at http://www.starlight.ufsc.br.

\section{results}\label{results}

\begin{figure*}
\begin{minipage}[b]{0.45\linewidth}
\includegraphics[width=\linewidth]{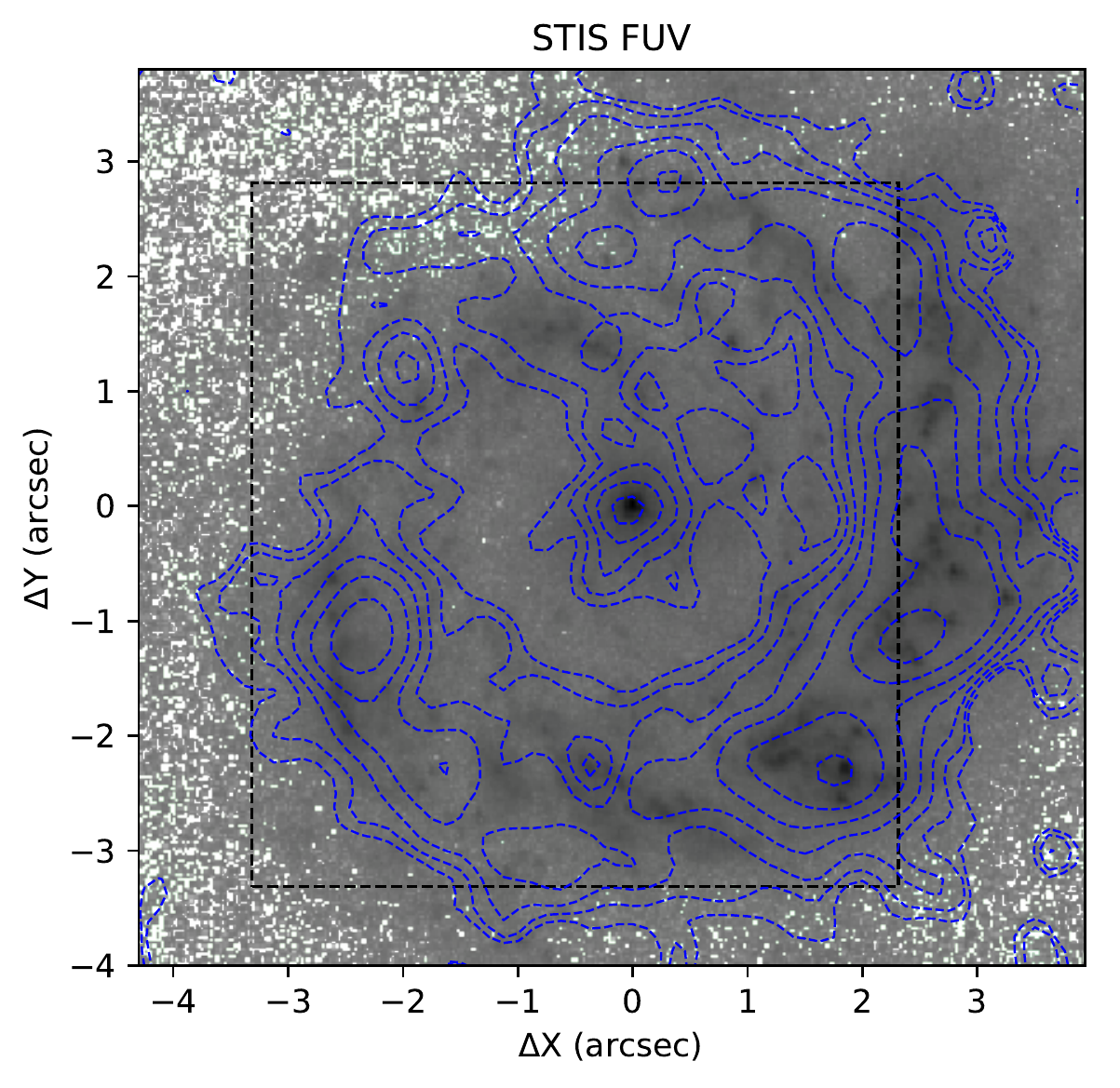}
\end{minipage} \hfill
\begin{minipage}[b]{0.50\linewidth}
\includegraphics[width=\linewidth]{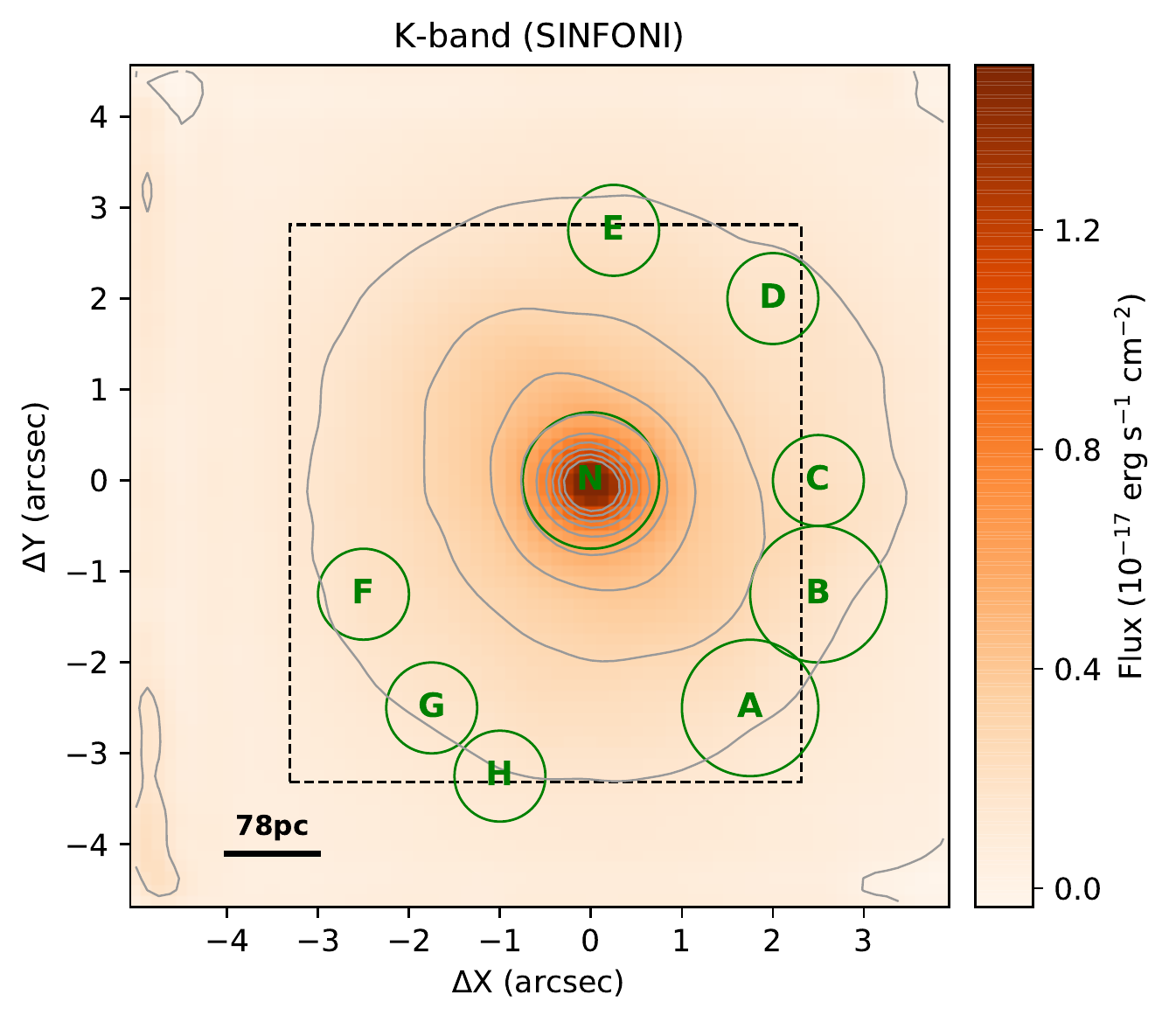}
\end{minipage} \hfill
\begin{minipage}[b]{1.0\linewidth}
\includegraphics[width=\linewidth]{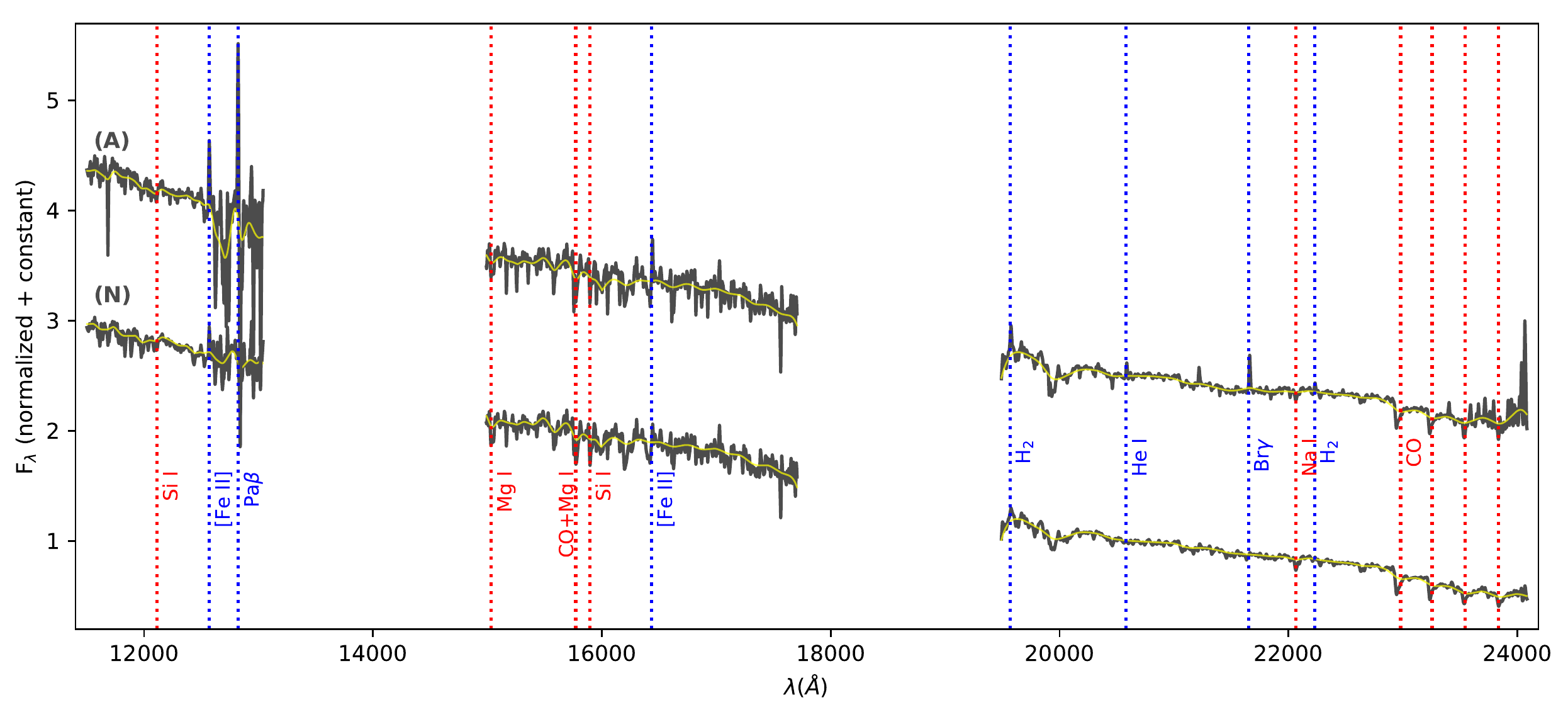}
\end{minipage}
\caption{STIS (Space Telescope Imaging Spectrograph) F25QTZ far ultraviolet image of NGC 4303 previously presented by \citetalias{colina02} in gray scale with Br$\gamma$ emission line (SINFONI) contours in blue (top left). K-band continuum (contours are shown in grey to help visualization), reconstructed from the SINFONI data cube as an average of the fluxes between 2.22 and 2.27 $\mu$m (top right). Green circles mark the position of the CNSFRs previously reported by \citetalias{riffelA16} and the black box in both images denotes the FoV used in this work. The bottom panel shows the near-IR spectra for the nucleus (bottom) and for position A (top), extracted with a circular aperture with radius 0\farcs75 and normalized at 2.067$\mu$m. The smoothed spectra are over-plotted in yellow.} A constant (1.5) was added to the spectrum of Region (A) for visualization purpose. Absorption (red) and emission (blue) lines are marked.
\label{ngc4303}
\end{figure*}

\begin{figure*}
\includegraphics[width=\linewidth]{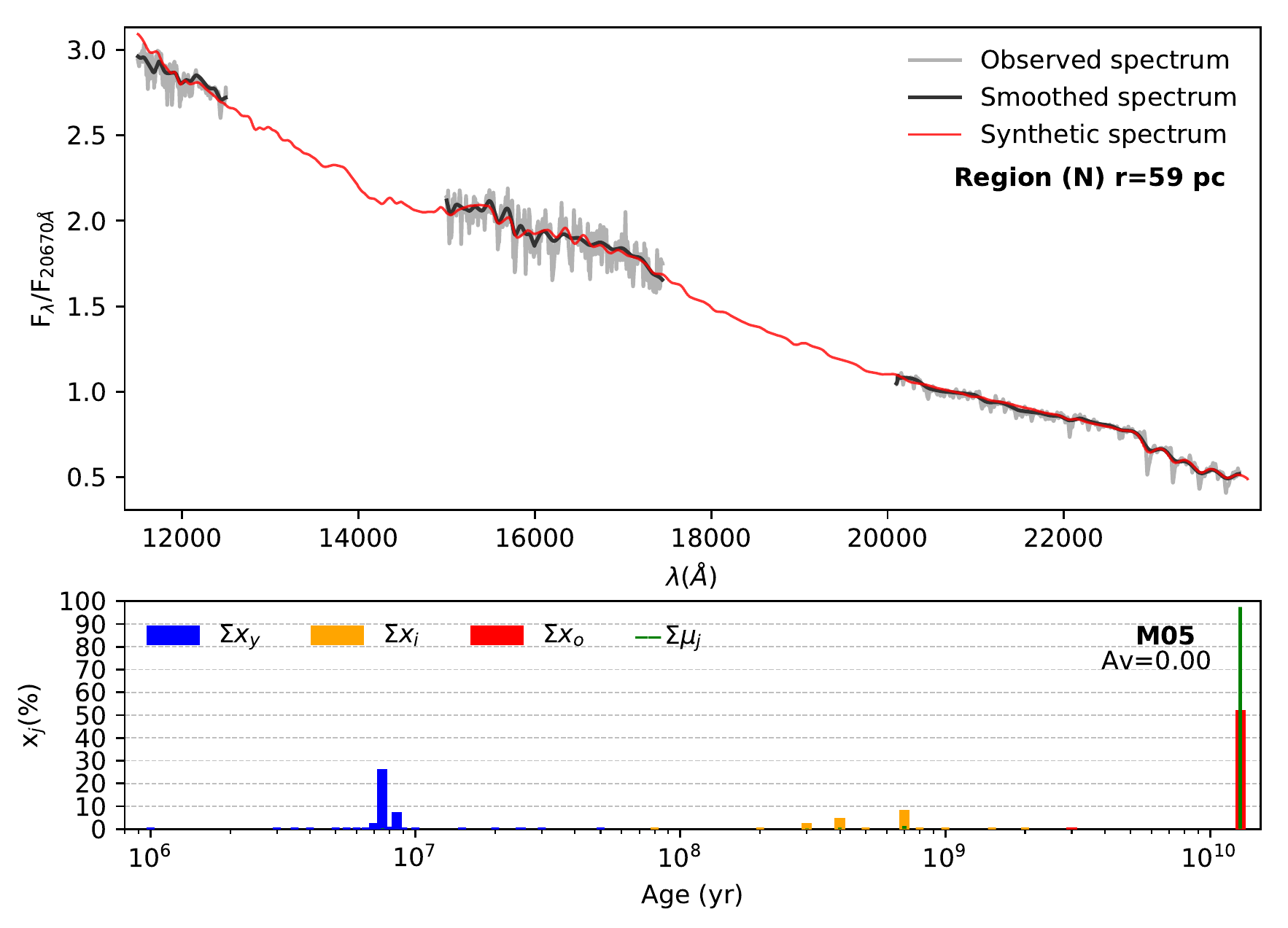}
\caption{Example of the stellar population synthesis results for the nuclear region (r=0\farcs75, corresponding to 60\,pc) of NGC\,4303 (central green circle in the maps). {\it Top panel}: observed (gray), smoothed (black) and synthetic (red) spectrum, normalized to unit at 2.067$\mu$m. {\it Bottom panel}: Histogram displaying the flux ($x_j$, colored bars) and mass-weighted (${\mu}_j$, green lines) stellar population vectors contributions sorted only by age (metallicities summed) and color coded by the three SPC age bins: young (blue: $\leq$ 50\,Myr), intermediate-age (orange: 0.05$-$2\,Gyr) and old (red: $>$ 2\,Gyr). The reddening value (in magnitude units) is also shown. Telluric absorption regions are omitted.}
\label{nuc}
\end{figure*}

Following \citet[][and references therein]{dametto14}, the stellar population vectors have been binned in three main components: \textit{young} (blue): $x_y$ ($t \leq 50\times10^6$ yr), \textit{intermediate-age} (orange): $x_i$ ($50\times10^6 < t \leq 2\times10^9$ yr) and \textit{old} (red): $x_o$ ($t > 2\times10^9$ yr). An example of the final fit for the nuclear region (r=0\farcs75$\sim$60\,pc, central green circle in Fig.~\ref{ngc4303}) of NGC\,4303 is presented in Fig.~\ref{nuc}. The nuclear spectrum is well described by a series of star formation bursts, the first one occurring $\sim$13\,Gyr ago\footnote{Note that the base is not a continuous distribution of ages, with the old ages being represented by 13\,Gyr SSPs in this case, for details see \citep{dametto14}.}, contributing with $\sim$50\% of the flux at 2.067$\mu$m. The fit reproduces individual minor bursts with ages ranging from 0.3\,Gyr to 0.7\,Gyr, corresponding to a 15\% contribution of the intermediate-age SPC, while the young SPC accounts for 35\% of the flux, with a major burst at 7.5\,Myr ago. No contribution of the featureless continuum and/or hot dust components were necessary in order to reproduce the nuclear continuum of this source. Green lines represent the percentage contribution in mass of each SSP. The old SPC ($m_o$) dominates the mass contribution with 98\%, while the contribution of the other two components are negligible. Moreover, the code fits a dust free ($A_V$=0.0) spectrum for the inner 60\,pc of NGC\,4303, agreeing with previous results from \citetalias{colina02},which found low extinction values (A$_V$=0.3\,mag) for the inner 0\farcs9 of this source.

\begin{figure*}
\begin{minipage}[b]{1.0\linewidth}
\includegraphics[width=\linewidth]{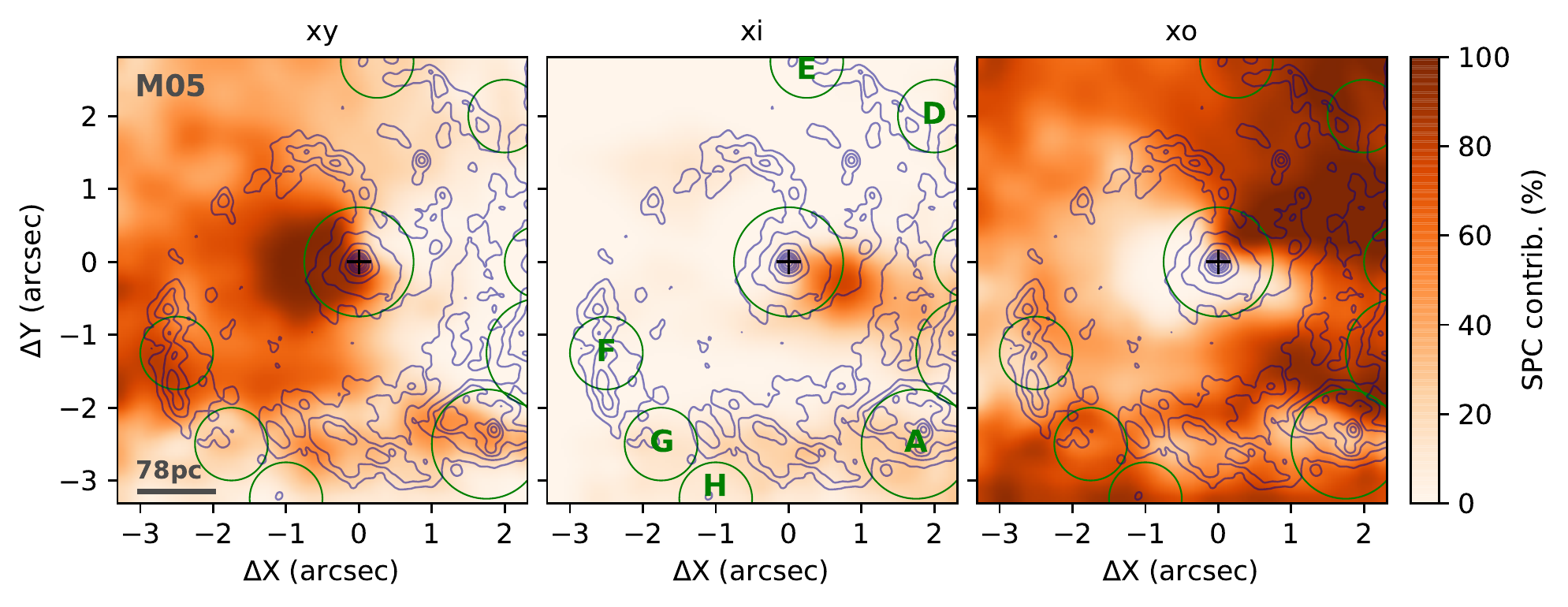}
\end{minipage} \hfill
\begin{minipage}[b]{1.0\linewidth}
\includegraphics[width=\linewidth]{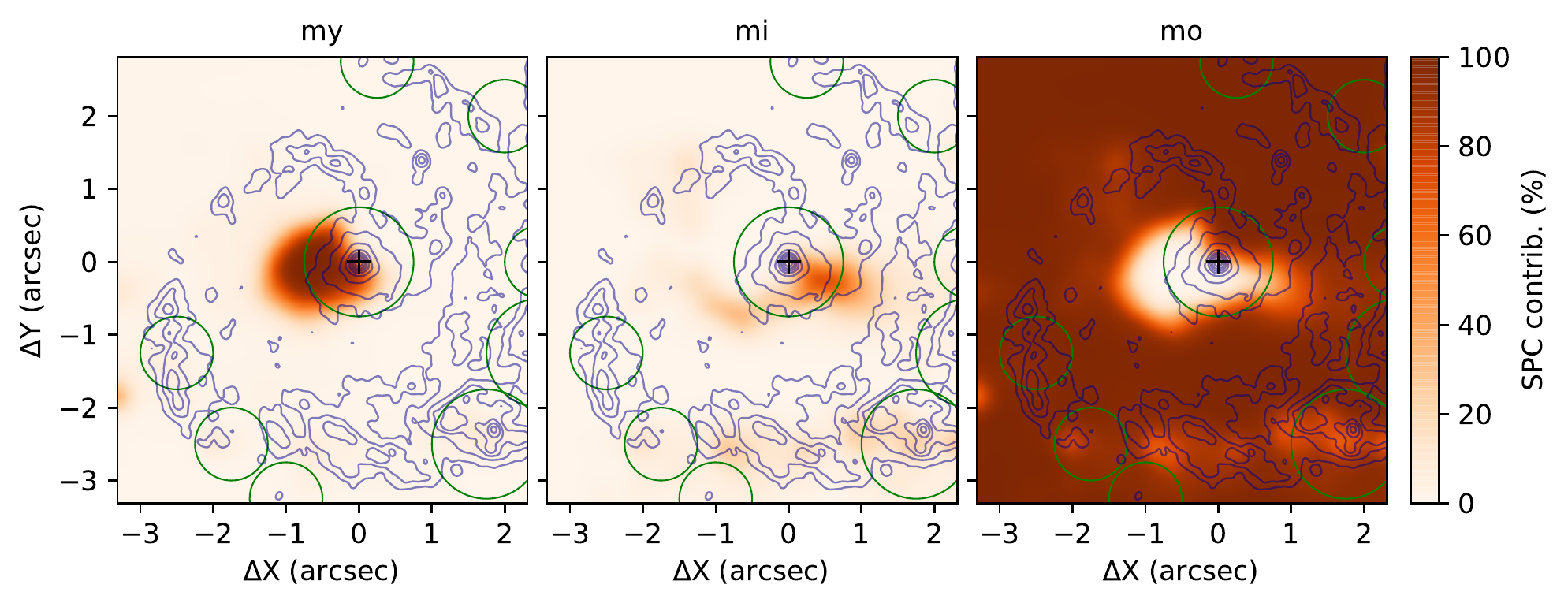}
\end{minipage}
\caption{Results using {\bf \citetalias{maraston05}} models. Spatial distribution of the percent contribution of each SPC to the flux ($x_j$, top panels) and mass ($m_j$, bottom panels), where $j$ represents the age of the SPC: young (y: $\leq$ 50\,Myr), intermediate-age (i: 0.05$-$2\,Gyr) and old (o: $>$ 2\,Gyr). Green circles mark the CNSFRs, clearly seen in the EW Br$\gamma$ previously reported by \citetalias{riffelA16}. The black cross represents the peak of the Br$\gamma$ emission line. STIS far-ultraviolet broad-band image is shown in blue contours, tracing the spiral structure reported in \citetalias{colina02}.}
\label{maps_sps}
\end{figure*}

Spatial distribution of the percent flux (top panels) and mass (bottom panels) contribution of each SPC bin is shown in Fig.~\ref{maps_sps}\footnote{We smoothed all maps to the H-band spatial resolution (0\arcsec.6). Note that the K-band has a higher (0\arcsec.5) spatial resolution, while the J-band has a lower one (1\arcsec). We decided to smooth the maps using the H-band resolution since we are using a small spectral region in the J-band to perform the fits.}. As for Fig.~\ref{ngc4303}, green circles represent the CNSFRs analyzed in \citetalias{riffelA16}. The physical sizes\footnote{The apertures were chosen by the authors in \citetalias{riffelA16} to be larger than the seeing ($\sim$0\farcs5) and to include most of the Br$\gamma$ flux of each region.} of these regions are: Regions N/A/B (r=0\farcs75$\sim$60\,pc); others (r=0\farcs5$\sim$39\,pc). We point out the FoV used in this work is smaller than that used to study the emission lines in \citetalias{riffelA16}, mainly due to problems with the relative flux calibration and  due to low SNR close to the borders, which represents a limitation when performing stellar population synthesis \citep{cid05a}. 

Analyzing the SPC contributions in Fig.~\ref{maps_sps}, we can see the age of the dominant stellar population presents a spatial variation, suggesting an age stratification along the inner $\sim$200\,pc radius of this source. Three main features are evident from the maps: A blob dominated by young stars in the nuclear region, shifted towards east from the centre (here defined as the peak of the Br$\gamma$ emission line) and the UV emission peak; a second blob dominated by intermediate-age stars located southwest from the centre, clearly seen in the middle panels; and a more internal arm-like structure very close to the blob of young stars, mainly seen in the $m_i$ map.

The youngest SPCs ($x_y$ and $x_i$) are distributed along the circumnuclear region (200$-$250 inner parsecs), similar to the results predicted from the emission line gas presented in \citetalias{riffelA16}. It is important to highlight that our FoV does not cover the whole western part of the circumnuclear ring (we miss most of regions {\it B} and {\it C}). In fact, a major contribution of the young SPC is seen in the northeastern area, which is co-spatial with dustier regions reported by \citepalias[][see their figure\,3]{colina00} using $V$-$H$ color. As we are using NIR data, we were able to penetrate deeper into the dust layers, accessing these young stars missed in the UV and optical ranges, which are more sensitive to dust obscuration. This could explain why the young SPC map is not tracing the spiral-arm structure clearly seen in the UV emission (Fig.~\ref{ngc4303}, darker knots represent less obscured regions), but the circumnuclear ring.  The old SPC is distributed outside the two blob structures, being more prominent northwest from the centre.

Flux-weighted results have a dependence on the choice of the normalization wavelength \citep{riffel11b}. Thus, one should take this into consideration when comparing results from different spectral regions. By using the knowledge of stellar evolution, one can use the mass-to-light ratio (M/L) for each SSP and determine the percentage contribution in stellar mass, a physical parameter which does not depend on the normalization wavelength used in the fit. The bottom panels of Fig.~\ref{maps_sps} present the mass-weighted contribution of the SPCs. As we can see, the major contribution in mass comes from old (t $> 2\,Gyr$) stars. In addition, the three internal structures mentioned above (the two blobs and the inner spiral arm) are highlighted in these maps.

\begin{figure*}
\begin{minipage}[b]{0.325\linewidth}
\includegraphics[width=\linewidth]{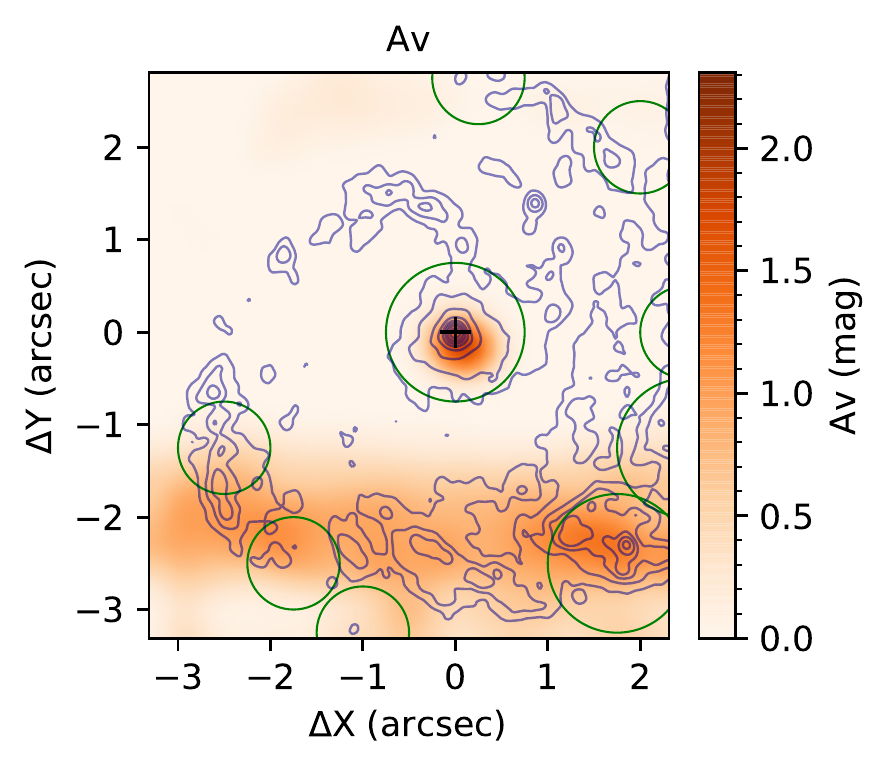}
\end{minipage} \hfill
\begin{minipage}[b]{0.335\linewidth}
\includegraphics[width=\linewidth]{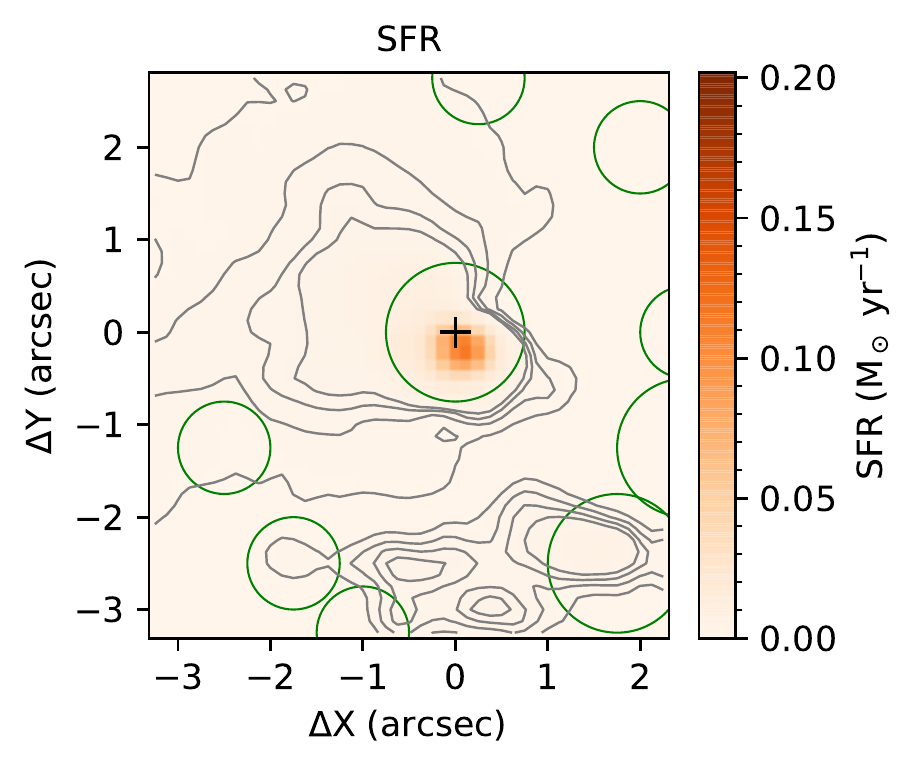}
\end{minipage}\hfill
\begin{minipage}[b]{0.32\linewidth}
\includegraphics[width=\linewidth]{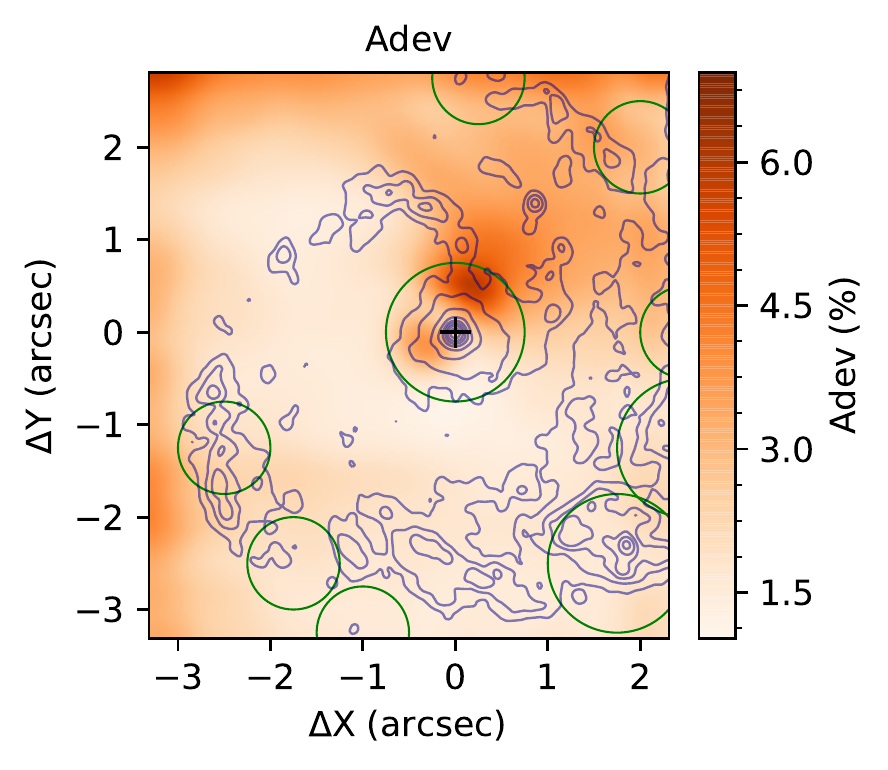}
\end{minipage}\hfill
\caption{{\it Left:} Reddening (A$_V$); {\it Middle:} Star Formation Rate (SFR) over the last 10\,Myr; {\it Right:} Adev (percent mean deviation from the spectral fit). Grey contours in the middle panel represent SFRs up to 2.5$\times$10$^{-3}$\,M$_{\odot}$yr$^{-1}$, were included for display purposes.}
\label{maps_extra}
\end{figure*}

Besides the SPC distributions, {\sc starlight} full spectral fitting provides a measure of the internal extinction (A$_V$), which is shown in the left panel of Fig.~\ref{maps_extra}. The nuclear spaxels clearly display the highest values (A$_V$=2.3\,mag), within scales of tens of parsecs, while the average value over the FoV is $\sim$0.3\,mag. The southern region of the reddening map displays a stripe-like pattern with reddening values around 1.5\,mag. This stripe-like pattern may be partially related with a residual `instrumental fingerprint', which we were not able to remove completely (see Sec.~\ref{fp}). Thus, possibly affecting the reddening values in these locations.


{\sc starlight} code also outputs the mass that has been processed into stars over the last t years (M$^t_{\star}$). This can be used to estimate the mean SFR over a period of time $t$. We have estimated the mean SFR$_{\star}$ over the last 10\,Myr as being the ratio of M$^t_{\star}$/$t$ (for t$\leq$10\,Myr), and it is presented in the middle panel of Fig.~\ref{maps_extra}. The nucleus clearly stands up with a SFR of $\sim$0.2\,M$_\odot$yr$^{-1}$ in the region co-spatial with the A$_V$ peak ($\sim$40\,pc radius). The mean SFR over the whole FoV is 2.1$^{+9.7}_{-2.1} \times$10$^{-3}$M$_\odot$yr$^{-1}$.

As mentioned in Sec.~\ref{method}, the quality of the fit can be measured by the {\sc starlight} output parameter called {\it percent mean deviation}: Adev ($|O_{\lambda} - M_{\lambda}|/O_{\lambda}$), in which $O_{\lambda}$ is the observed spectrum and $M_{\lambda}$ is the fitted model \citep{cid04,cid05b}. For our fits, Adev is below 4.5\% at most locations (see Fig.~\ref{maps_extra}, right panel), indicating the model reproduces well the observed spectra. 

\subsection{Robustness of the stellar population results: Comparison with BC03 and C09 SSP models}\label{mc}

In order to deepen our analysis, we decided to perform stellar population synthesis using other EPS models available in the literature: \citet[][hereafter BC03]{bc03} and \citet[][hereafter C09]{conroy09}. We do not include the models from \citet[][hereafter M11]{maraston11} in this analysis, as the use of the lower resolution \citetalias{maraston05} models is a better option to deal with stellar population synthesis in the NIR when compared to the M11 ones \citep{dametto14}. Addressing to the limitations on the \citetalias{maraston11} models, an example is the fact that above $\sim$1$\mu$m about half of the spectra \citep[from Pickles stellar spectral library, used to construct the \citetalias{maraston11} models that extent to the NIR,][]{pick98} lack spectroscopic observations leading the authors to construct a smooth energy distribution from broad-band photometry, which may imply that some NIR absorption features are not well resolved, even for these higher resolution models. New sets of models presented by \citet{meneses15}, \citet{rock16} and \citet{conroy18} are available, which make use of the IRTF and E-IRTF stellar libraries. However, the age range of these models (t $\geq$ 1\,Gyr) is not adequate to fit the spectrum of galaxies with active star formation, therefore these models were not employed here. 

Firstly, the three models used in this work are constructed using different prescriptions \citep[see][for a review]{conroy13}. Both \citetalias{bc03} and \citetalias{conroy09} models are constructed using the isochrone synthesis approach, while \citetalias{maraston05} uses the fuel consumption theory technique. In the former, the SSPs are constructed by integrating the contributions of all mass bins (along one isochrone) to the flux in the various passbands, after assuming a initial mass function (IMF). Yet, in the later, energetics of the post-main-sequence phases (i.e. the amount of fuel available for nuclear burning) are calculated using the evolutionary track of the turnoff mass.

The treatment of the TP-AGB phase - crucial to models the stellar populations in the NIR - is also a topic of discussion when comparing the results obtained with different EPS models. While \citetalias{bc03} constructed period-averaged spectra for C-type stars using broad-band photometry to calibrate the low-resolution stellar templates of \citet{hofner00}, both \citetalias{maraston05} and \citetalias{conroy09} include empirical spectra of carbon- and oxygen-rich stars from \citet{lancon02}. The inclusion of these empirical spectra of stars in the TP-AGB phase has enabled the detection of NIR characteristic absorption features, such as TiO (0.843 and 0.886$\mu$m), VO (1.048$\mu$m), CN (1.1 and 1.4$\mu$m) and CO (1.6 and 2.3 mum) bands \citep{riffel15}. 

In recent studies, there has been an attempt to address discrepancies in stellar population synthesis results using different EPS models. \citet{baldwin18} studying a sample of 12 nearby early-type galaxies recently concluded the variation in the derived SFHs using NIR is largely driven by the choice of stellar spectral library rather than the models' prescription. In order to better compare the results, our base sets are composed by SSPs constructed using basically the same stellar library (BaSeL\footnote{\citetalias{maraston05}: BaSeL\,2.2 \citep{lejeune97,lejeune98} and \citetalias{bc03}/\citetalias{conroy09}: BaSeL\,3.1 \citep{lejeune97,lejeune98,westera02}}) in the NIR spectral range. 

Performing stellar population synthesis with STARLIGHT code for 7 spectra from early-type galaxies, \citet{dahmer18} found systematic differences in the results using \citetalias{maraston05}, \citetalias{bc03} and \citetalias{conroy09} models. While \citetalias{bc03} presented a higher contribution of young stellar populations, \citetalias{conroy09} displayed a major contribution of the older ages and \citetalias{maraston05} preferred solutions including a higher contribution (when compared to the other two models) of the intermediate-age components. One way to diminish such systematic effects is to fix the kinematic fit while running the code. As we degrade the resolution of the data to match the low resolution of these models in the NIR, the stellar features are considerably broadened, thus one cannot rely on the kinematic information obtained with STARLIGHT. 

We have tested the results keeping the kinematics parameters fixed for NGC\,4303 (see Sec.~\ref{code}) and we were able to smooth the systematic effects found by \citet{dahmer18}, concluding the inclusion of a kinematic fit while running STARLIGHT with low resolution models (R$\sim$300) do not yield reliable results, since the stellar features in the data are considerably broadened.

\begin{figure*}
\includegraphics[width=\linewidth]{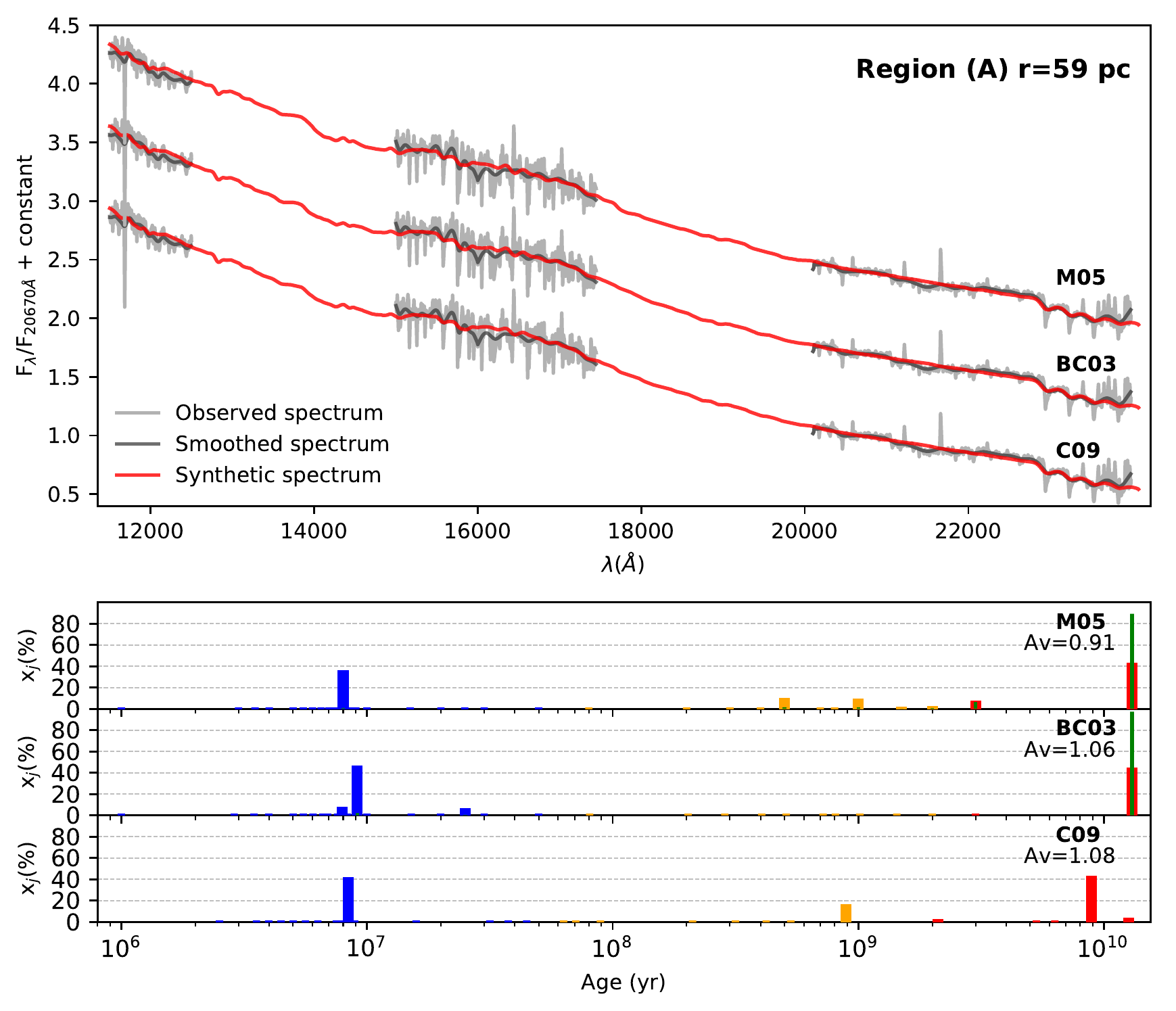}
\caption{Same as Fig.~\ref{nuc}, but for {\it Region\,A} (see Fig.~\ref{ngc4303}). Results using different EPS models are presented.}
\label{nuc_models}
\end{figure*}

\begin{figure*}
\begin{minipage}[b]{1.0\linewidth}
\includegraphics[width=\linewidth]{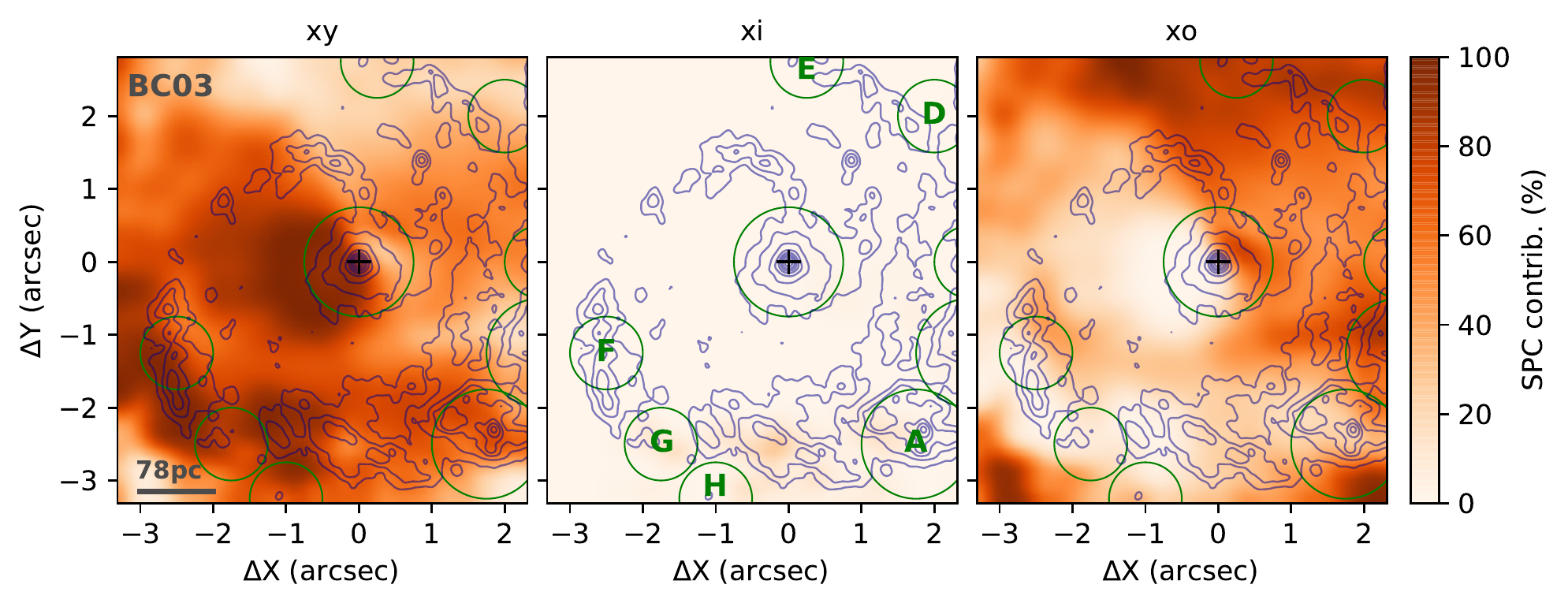}
\end{minipage} \hfill
\begin{minipage}[b]{1.0\linewidth}
\includegraphics[width=\linewidth]{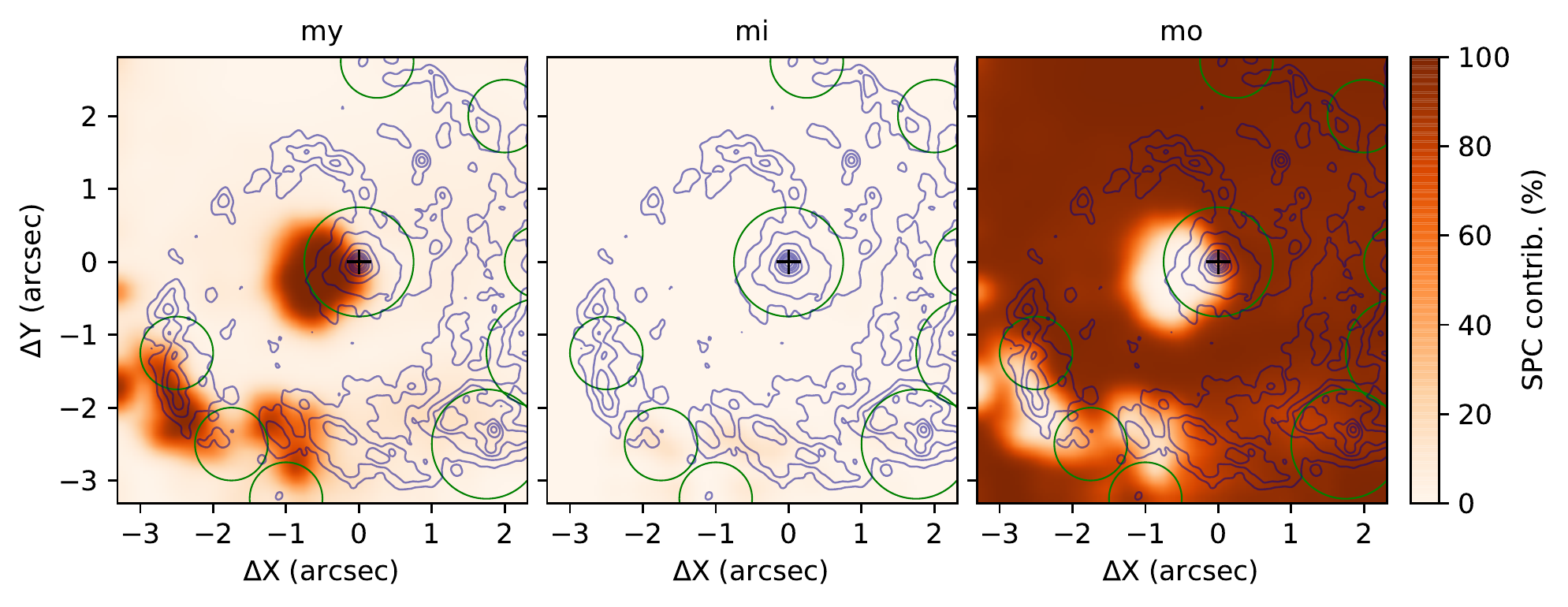}
\end{minipage}
\caption{Same as Fig.~\ref{maps_sps}, but using {\bf BC03} models.}
\label{maps_sps_BC03}
\end{figure*}

\begin{figure*}
\includegraphics[width=\linewidth]{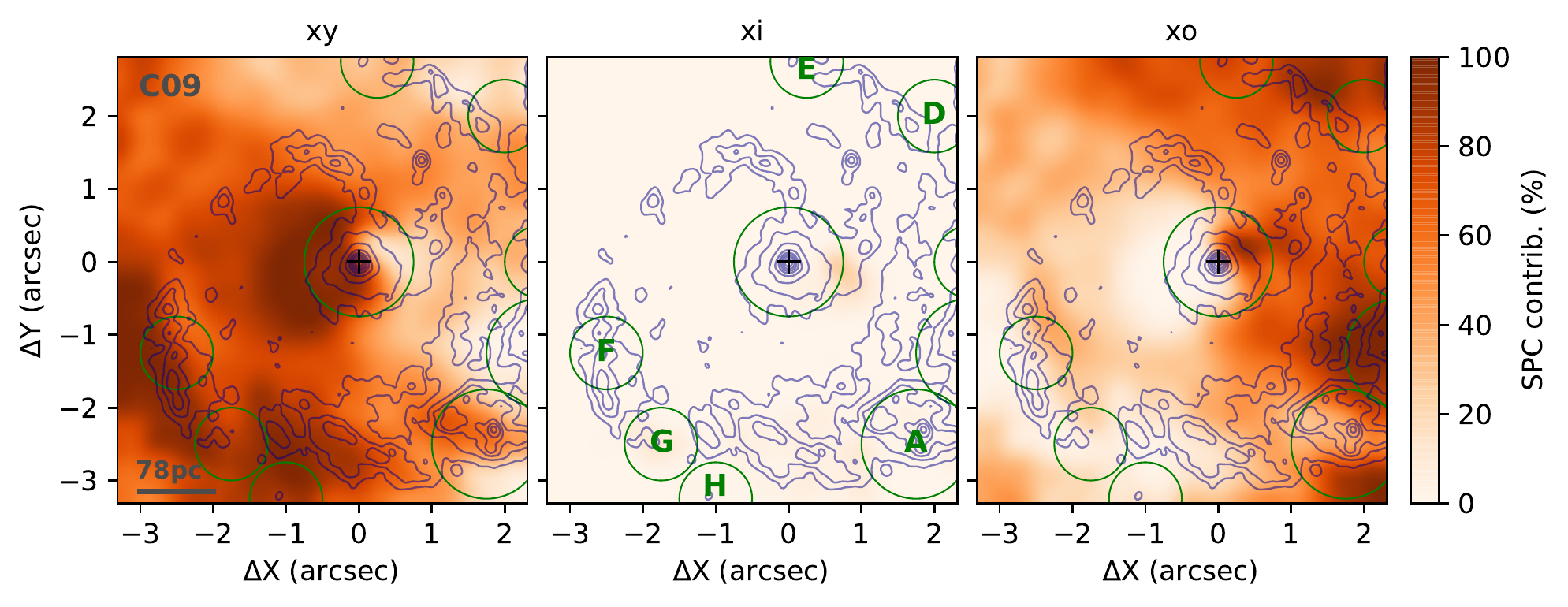}
\caption{Same as Fig.~\ref{maps_sps}, but using {\bf \citetalias{conroy09}} models. As these models do not provide the mass-to-light ratio needed to calculate the mass-weighted contributions, we only present the maps for the flux-weighted contribution of each SPC.}
\label{maps_sps_C09}
\end{figure*}

Base sets for \citetalias{bc03} and \citetalias{conroy09} were constructed in the same way as that of \citetalias{maraston05} and the same fitting procedure was applied (see Figs~\ref{nuc_models} to \ref{maps_sps_C09}). \citetalias{conroy09} models do not provide the fraction of the initial stellar mass which is still present in form of stars at the age $j$ for each base set component.Therefore we were not able to calculate the percentage mass contribution using these models and we only present the flux-weighted results.

In Fig.~\ref{nuc_models} we present an example of the fits for {\it Region A} (see Fig.~\ref{ngc4303}) using the three EPS models for comparison. From the top panel we can see the overall good quality of the fits, primarily in the K-band. Analyzing the histograms in the bottom panels, it is clear the results assuming the binned stellar population vectors are in agreement between the models, presenting a $\sim$40\% contribution for both young (8$-$9\,Gyr) and old (9/13\,Gyr) SPC, while the remaining $\sim$20\% comes from intermediate-age stars (0.5$-$1\,Gyr), with exception of \citetalias{bc03} results which do not display any contribution of the $x_i$ component and present an increase in $x_y$ ($\sim$60\%).

The reddening solutions are similar, deviating at most in 0.17\,mag, which can be explained by the fact that the NIR spectral range is less sensitive to reddening variations when using full spectrum fitting \citep{baldwin18,dahmer18}.

In Figs~\ref{maps_sps_BC03} and \ref{maps_sps_C09} we present the spatial distribution of the SPCs using \citetalias{bc03} and \citetalias{conroy09} models. The same trend of age stratification found using \citetalias{maraston05} models is reproduced in these figures: The young SPC is distributed along the circumnuclear region with clear knots shifted eastwards from the photometric center and the spiral-arm structure found in the UV (blue contours); the old SPC is distributed outside these knots, with a clear increase in the contribution towards northwest; the intermediate-age contribution is almost negligible using these sets of models, with small knots in the south region for \citetalias{bc03} and next to the centre (co-spatial with the intermediate-age blob found using \citetalias{maraston05} models) for \citetalias{conroy09}.

\begin{figure*}
\begin{minipage}[b]{1.0\linewidth}
\includegraphics[width=\linewidth]{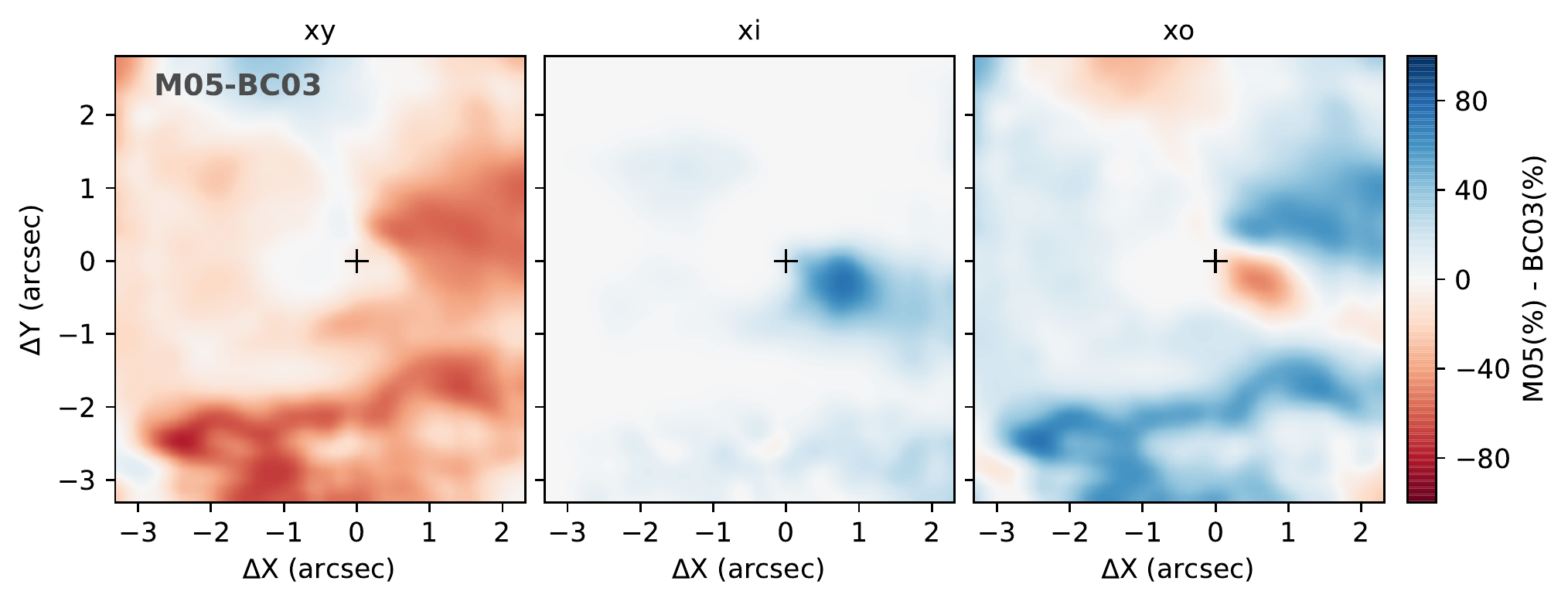}
\end{minipage} \hfill
\begin{minipage}[b]{1.0\linewidth}
\includegraphics[width=\linewidth]{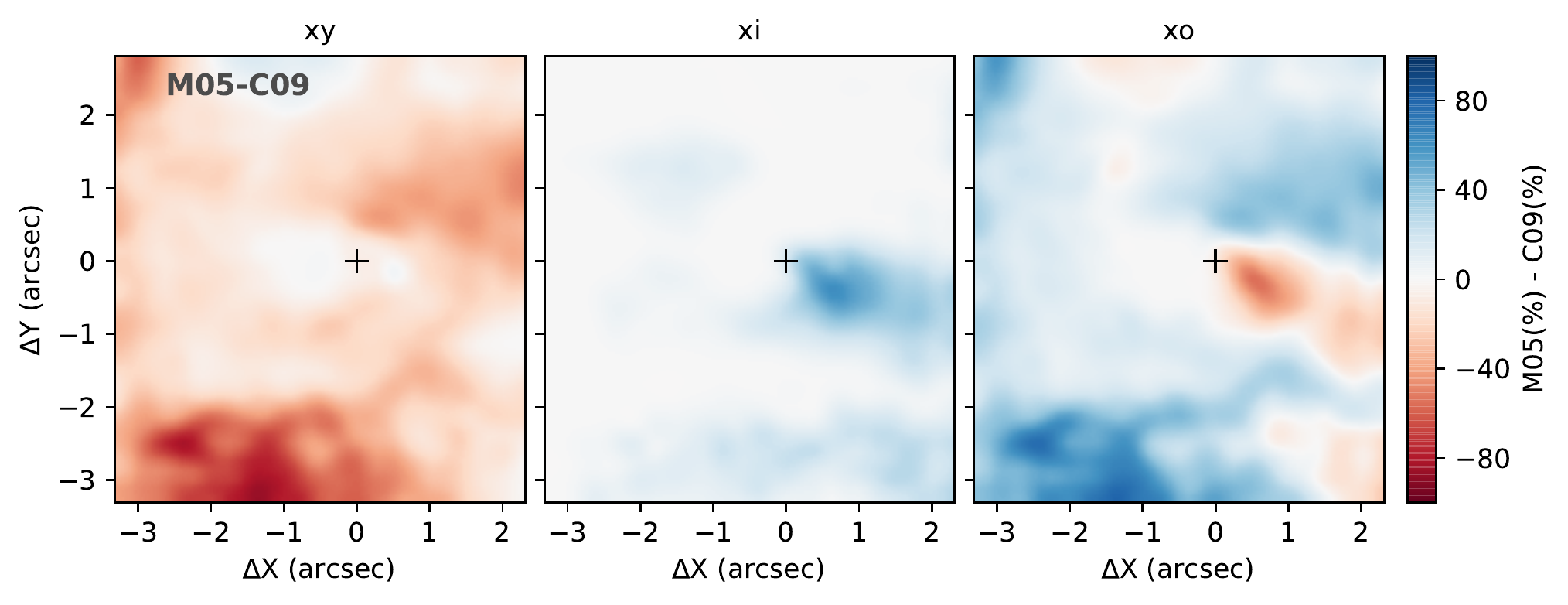}
\end{minipage} \hfill
\begin{minipage}[b]{1.0\linewidth}
\includegraphics[width=\linewidth]{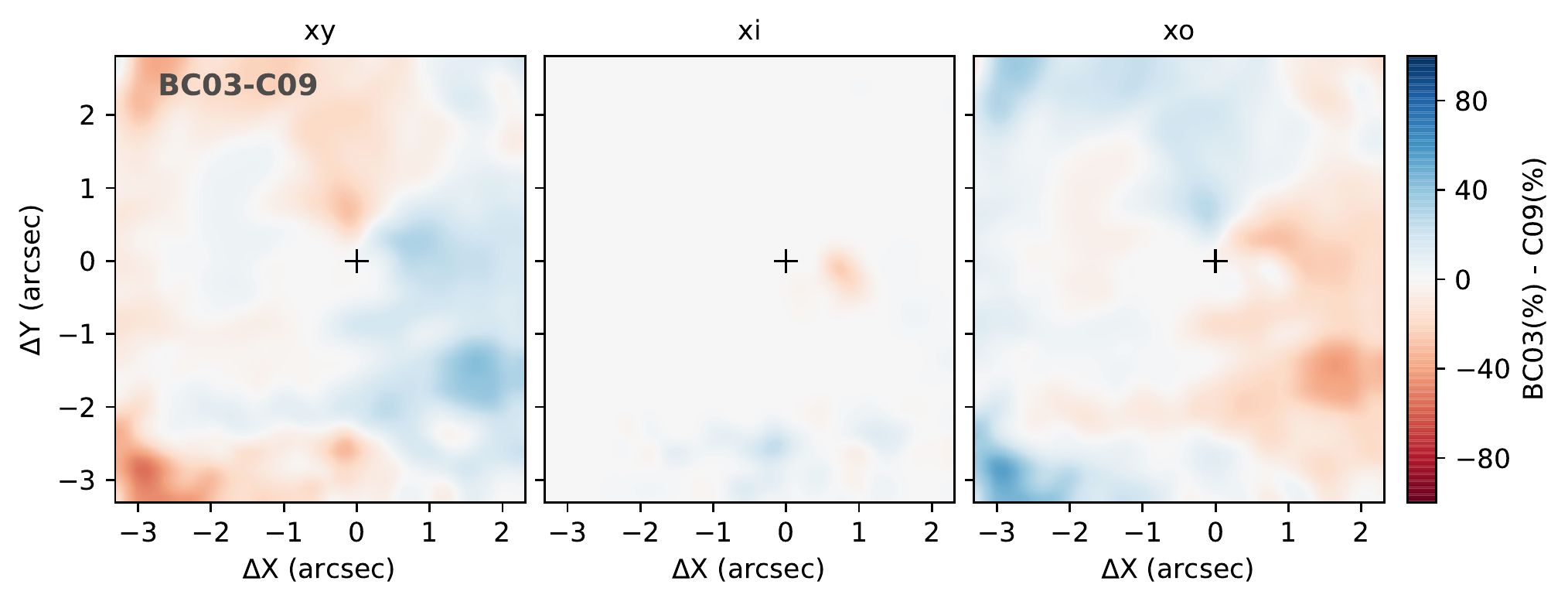}
\end{minipage}
\caption{Comparison between the results produced with the different EPS models. Maps displaying the difference between the flux-weighted SPCs presented in Fig.~\ref{maps_sps}, \ref{maps_sps_BC03} and \ref{maps_sps_C09}.}
\label{maps_compare}
\end{figure*}

In order to better compare the results, we present maps with the spatially resolved differences in the SPC vectors (flux-weighted) between the three models in Fig.~\ref{maps_compare}. From the two first columns, we can see \citetalias{bc03} and \citetalias{conroy09} display higher contribution of the $x_y$ SPC (redder colors in the left panels), while \citetalias{maraston05} favors the $x_i$ and $x_o$ SPCs (bluer colors in the middle and right panels). Major differences are found in the southern region, co-spatial with the residual 'instrumental fingerprint' from SINFONI data (see Sec.~\ref{fp}), which may be the cause of these discrepancies in this region. Moreover, the intermediate-age blob found using \citetalias{maraston05} models appears to be spread in the older ages using \citetalias{bc03} and \citetalias{conroy09} models, while the inner spiral-arm clearly seen in the mass-weighted \citetalias{maraston05} maps (Fig.~\ref{maps_sps}) and traced by bluer regions in the middle panels of Fig.~\ref{maps_compare} is missed using \citetalias{bc03} and \citetalias{conroy09} models. 

The smallest differences are displayed between \citetalias{bc03} and \citetalias{conroy09} models, as can be seen from the bottom column in Fig~\ref{maps_compare}. These two models also present nearly negligible contribution of the $x_i$ SPC, in contrast to results produced using \citetalias{maraston05} models. This could be related to the different prescription plus treatment of the TP-AGB phase for the different models. For example, \citet{baldwin18} argue their results favor \citet{maraston11} treatment of the TP-AGB phase \citepalias[same used in][]{maraston05}, in agreement with \citet{riffel15}. \citet{baldwin18} also claim that the discrepancies found by \citet{zibetti13}, which states \citetalias{maraston05} models overestimate the TP-AGB contribution, are probably related to the technique used in \citet{zibetti13}, consisting of measuring line indices rather than full-spectrum fitting.

We decided to discuss only the results for \citetalias{maraston05} in Sec.\ref{discussion}, since these models present a proper treatment of the TP-AGB phase as discussed above, as well as the informations needed to calculate the percentage mass contributions of the stellar populations. Moreover, this will allow us to compare the results presented here with those previously published by our group using \citetalias{maraston05} models \citep[e.g.][]{riffelA10,riffel11c,storchi12,dametto14,schonell17,diniz17}.

\section{discussion}\label{discussion}

\subsection{Distribution of the stellar populations: circumnuclear ring and spiral structures}

\begin{figure}
\begin{minipage}[b]{1.0\linewidth}
\includegraphics[width=\linewidth]{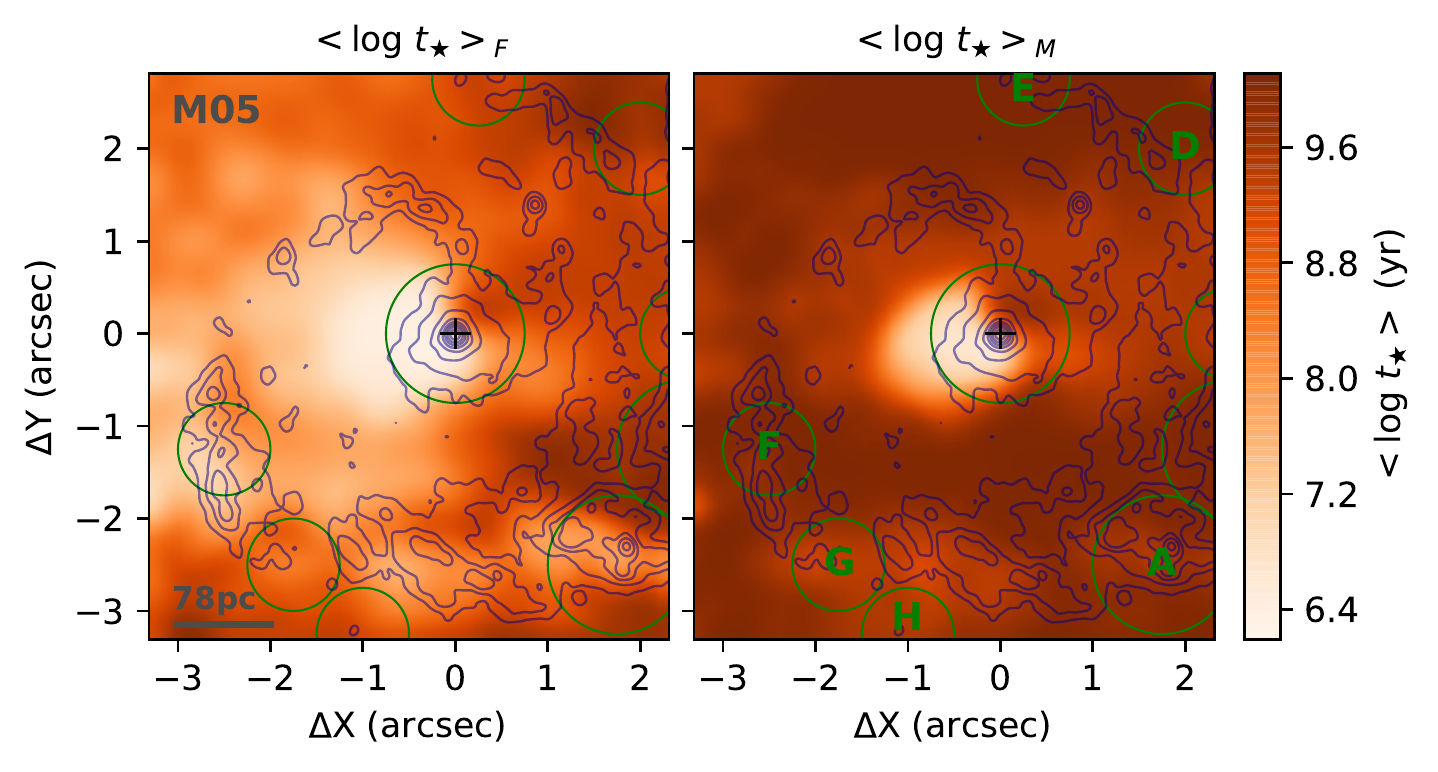}
\end{minipage} \hfill
\begin{minipage}[b]{1.0\linewidth}
\includegraphics[width=\linewidth]{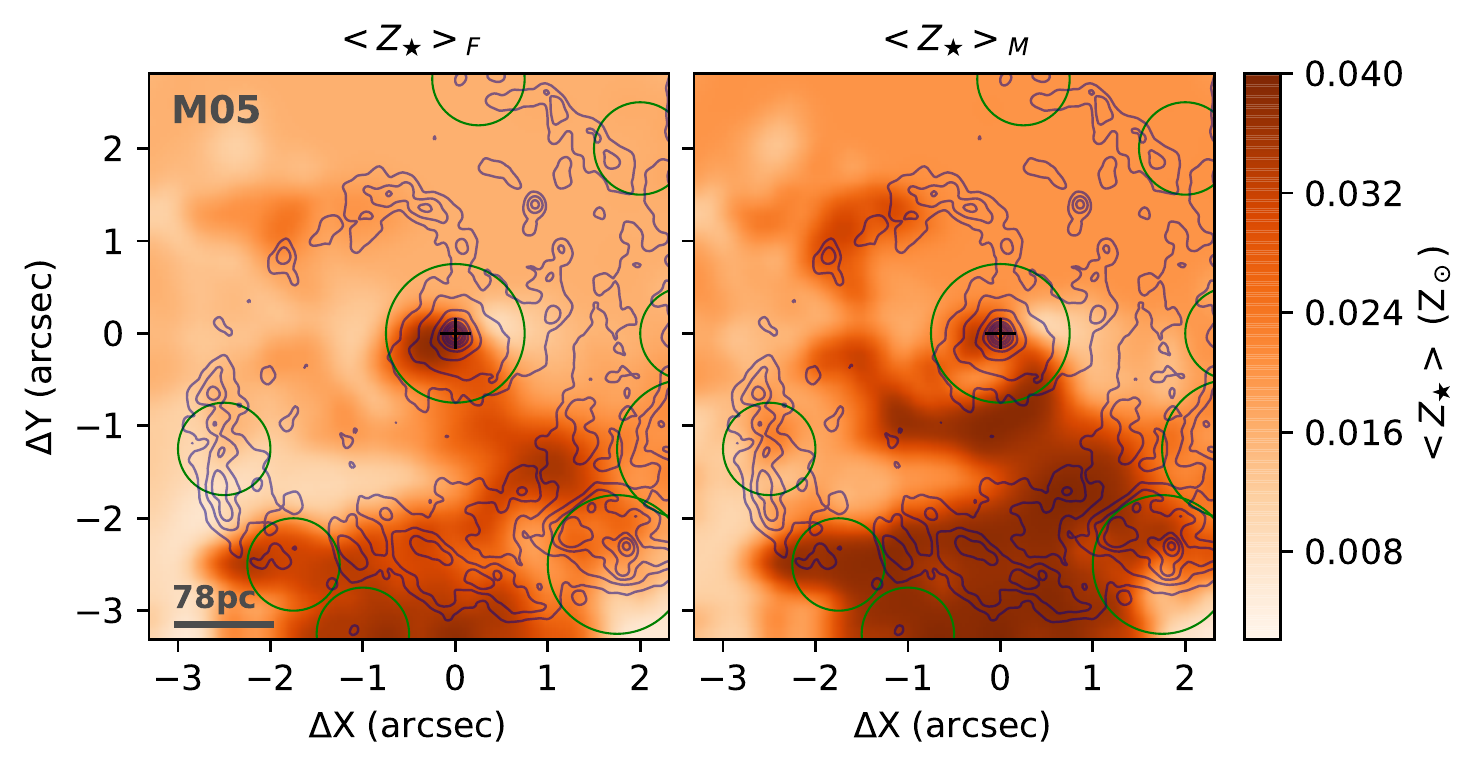}
\end{minipage}
\caption{Logarithm of the mean age (top panels) and mean metallicity (bottom panels) weighted by flux ({\it left}) and weighted by mass ({\it right}). Black cross represents the centre, blue contours from UV image and green circles marking the CNSFRs (see Fig.~\ref{ngc4303} for more details).}
\label{maps_mean}
\end{figure}

Stellar population distribution in the inner $\sim$200\,pc radius of NGC\,4303 shows the presence of young and intermediate-age stars in a circumnuclear ring structure (see the mean age maps presented in Fig.~\ref{maps_mean}), while the older components are distributed mainly along the more external parts of our FoV and we interpret them as the underlying bulge component. Previous 2D mapping studies of our AGNIFS group on Seyfert galaxies showed a predominance of intermediate-age stars in the central kiloparsecs of these objects \citep{riffelA10,riffel11c,storchi12,schonell17}, while our results point to a more conspicuous contribution of this SPC in the LLAGN NGC\,4303: a blob southwest from the centre which extends to an internal spiral-arm like structure surrounding the central blob of young stars. These features suggest star formation is stratified along the inner regions of NGC\,4303, with older knots shifted towards west of the centre, while the younger ones are to the east. \citetalias{colina02} and \citetalias{riffelA16} reported younger ages for the western knots, which are missed by our smaller FoV, it is important to highlight. 

As discussed in \citetalias{riffelA16}, our results favor the interpretation of a circumnuclear ring with young/intermediate-age stars rather than a spiral-arm structure proposed by \citetalias{colina02}. We suggest the spiral structure traced by the UV emission is in fact tracing a region of low extinction inside the disk containing the circumnuclear ring. This interpretation is in agreement with previous results from \citetalias{colina00}. These authors presented a {\it V$-$H} image of the central region of this source (see their figure~3), which reveled a rather complex gas/dust and stellar distribution with two-arm (star forming lane in the southwest and west regions, and dust lane at northeast) spiral structure. 

In the aim of better visualizing these findings we constructed mean value maps. A more compressed but also useful way to represent the stellar population mixture in the galaxy is by computing the mean stellar age (flux and mass-weighted, respectively), as defined by \citet{cid05b}:
\begin{equation}
\left< \log t_{\star} \right>_F = \sum_{j=1}^{N_{\star}} x_j log (t_j),
\end{equation}.
\begin{equation}
\left< \log t_{\star} \right>_M = \sum_{j=1}^{N_{\star}} m_j log (t_j).
\end{equation}
\noindent and mean stellar metallicity:
\begin{equation}
\left< Z_{\star} \right>_F = \sum_{j=1}^{N^{\star}} x_j Z_j,
\end{equation}
\begin{equation}
\left< Z_{\star} \right>_M = \sum_{j=1}^{N^{\star}} m_j Z_j.
\end{equation}
\noindent We present the maps for these parameters in Fig.~\ref{maps_mean}. From the mean age maps we can see the youngest stellar populations are co-spatial with the UV knots traced by the blue contours as well as the dustier regions at north-east probed by optical imaging, as mentioned before. A comparison between the mean age and mean metallicity maps suggests the presence of a rather young metal rich SPC ($\sim$2Z$\odot$) in the nuclear region of NGC\,4303, as well as in the southern region, where the youngest CNSFRs encompassed by our FoV are located. Moreover, higher metallicity values (up to 2\,Z$\odot$) are tracing the inner spiral-arm structure of intermediate-age stars (see also Fig.~\ref{maps_sps}). These results give further support to the age stratification scenario proposed for the SFH in this source. 

\subsection{Stellar populations in the CNSFRs}\label{cnsfrs}

In \citetalias{riffelA16} we present new maps of emission-line flux distributions and kinematics in both ionized and molecular gas in the inner 350\,pc radius (a slightly wider FoV than the one we use in this work, see Sec.~\ref{results}) of NGC 4303. The most prominent feature is a 200$-$250\,pc ring of CNSFRs which is seen as a nuclear spiral in UV/HST images presented in \citetalias{colina02}. In order to better analyze and compare our results with those obtained from the study of the emission-lines in \citetalias{riffelA16}, we decided to extract the spectra of the CNSFRs encompassed by our FoV, matching the aperture defined in that work. The spectra were extracted as the summed flux inside each aperture (Regions: N/A, r=60\,pc; D/E/F/G/H, r=39\,pc).

\begin{figure}
\includegraphics[width=\linewidth]{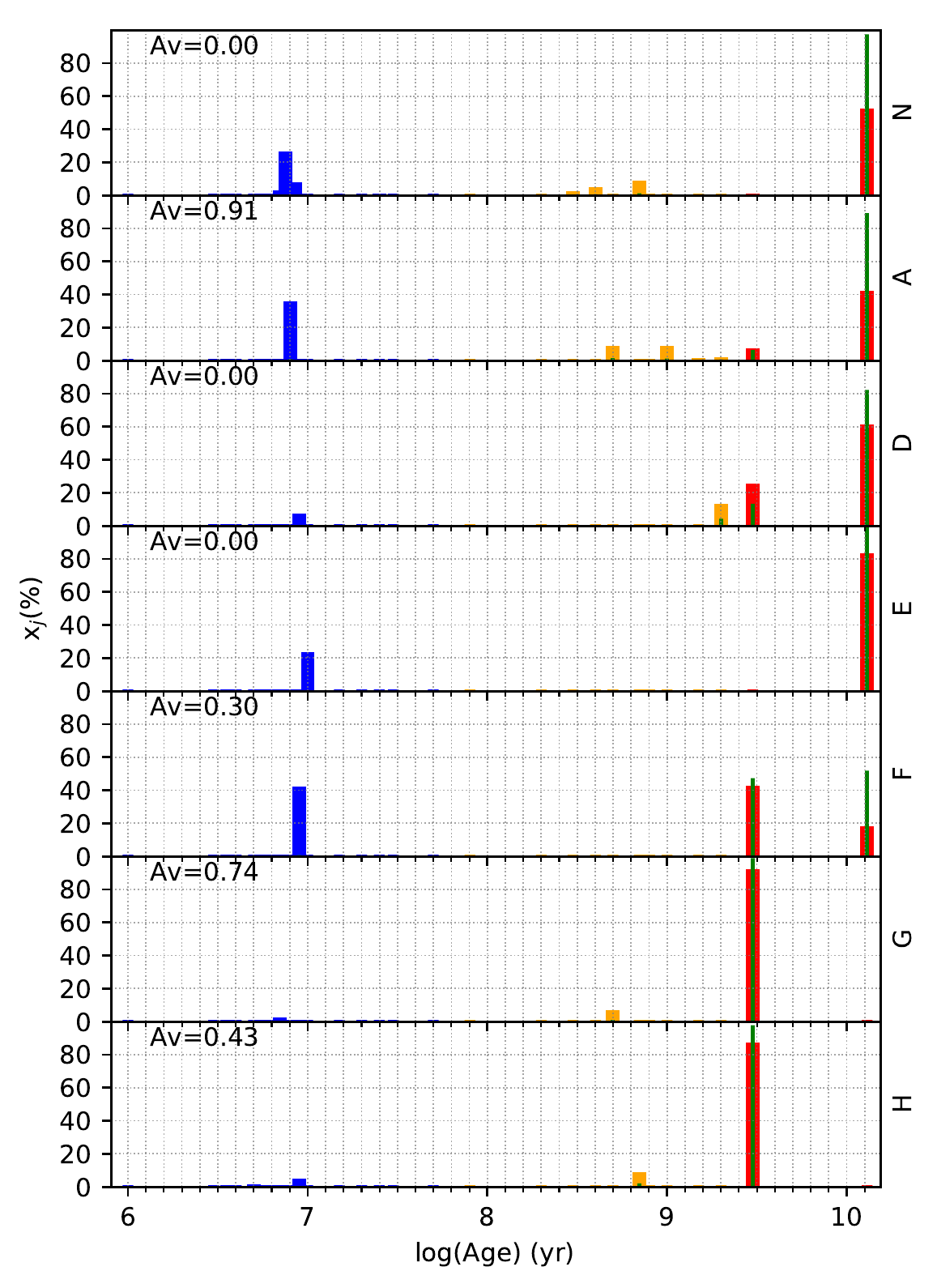}
\caption{SP synthesis results for the CNSFRs in NGC\,4303. Each panel displays the percent contribution in flux (color bars) and mass (green lines) at $\lambda$ = 2.067$\mu$m of the stellar population vectors in each region. Blue, orange and red represent young ($x_y$), intermediate-age ($x_i$) and old ($x_o$) SPCs contributions respectively; The letters correspond to the regions identified by \citetalias{riffelA16} and marked as green circles in our maps.}
\label{reg_m05}
\end{figure}

\begin{table*}
\centering
\scriptsize
\caption{Synthesis results for the CNSFRs shown in Fig.~\ref{reg_m05} (green circles in the maps).}
\label{sr}
\begin{tabular}{llllllllllllllll}
\noalign{\smallskip}
\hline
\noalign{\smallskip}
Region & {\sc $x_y$}  & {\sc $x_i$}  & {\sc $x_o$}  & {\sc $m_y$}  & {\sc $m_i$}  & {\sc $m_o$}  & $< \log$ $t_{\star} >_F$ & $< \log$ $t_{\star} >_M$ & {\sc Z$_F$} & {\sc Z$_M$} & A$_V$ & SFR  & $\sum_{SFR}$ & Adev \\
& (\%)  & (\%)  & (\%)  & (\%)  & (\%)  & (\%) & (yr) & (yr) & ({\sc Z$_{\odot}$}) & ({\sc Z$_{\odot}$}) & (mag) & (M$_\odot$yr$^{-1}$) & (M$_\odot$yr$^{-1}$kpc$^{-2}$) & (\%) \\
\noalign{\smallskip}
\hline
{\bf M05}  &   {\it (1)}   &  {\it (2)}    &  {\it (3)}    &   {\it (4)}    & {\it (5)}      & {\it (6)}      &  {\it (7)}    &  {\it (8)}      & {\it (9)}       & {\it (10)}      & {\it (11)} & {\it (12)} & {\it (13)} & {\it (14)}\\
\noalign{\smallskip}
(N)  &  35   &  14  &  51  &   0  &   2  &   97  & 6.17$\times$10$^8$ & 1.74$\times$10$^{10}$ & 1.81  &  1.98 &    0   &  0.426  & 39.62 &  1.04   \\
(A)  &  35   &  18  &  47  &   1  &   4  &   96  & 5.50$\times$10$^8$ & 1.00$\times$10$^{10}$ & 1.20  &  1.83 &  0.91  &  0.144  & 13.39 &  1.65   \\
(D)  &  6    &  12  &  82  &   0  &   4  &   96  & 4.68$\times$10$^9$ & 9.77$\times$10$^9$ & 1.00  &  1.00 &    0   &  0.005  & 1.0   &  2.90   \\
(E)  &  22   &  0   &  78  &   0  &   0  &   100 & 2.75$\times$10$^9$ & 1.29$\times$10$^9$ & 1.00  &  1.00 &    0   &  0.018  & 3.66  &  3.35   \\
(F)  &  41   &  0   &  59  &   1  &   0  &   99  & 3.63$\times$10$^8$ & 6.17$\times$10$^9$ & 1.33  &  1.21 &  0.30  &  0.046  & 9.71  &  1.98   \\
(G)  &  2    &  6   &  92  &   0  &   1  &   99  & 2.45$\times$10$^9$ & 2.95$\times$10$^9$ & 1.88  &  1.98 &  0.74  &  0.002  & 0.46  &  1.87   \\
(H)  &  5    &  8   &  87  &   0  &   2  &   98  & 1.95$\times$10$^9$ & 2.88$\times$10$^9$ & 1.85  &  1.96 &  0.43  &  0.005  & 1.09  &  1.53   \\
\noalign{\smallskip}
\hline
\noalign{\smallskip}
\end{tabular}
\begin{list}{Notes:}
\item (1), (2), (3): average contribution in flux of the young (y: $\leq$ 50\,Myr), intermediate-age (i: 0.05$-$2\,Gyr) and old (o: $>$ 2\,Gyr) SP component, respectively; (4), (5), (6): average contribution of the SP components in mass; (7), (8): flux- and mass-weighted mean ages; (9), (10): flux- and mass-weighted mean metallicities; (11): visual extinction; (12): star formation rate; (13): star formation surface density; (14): percent mean deviation.
\end{list}
\end{table*}

In Fig.~\ref{reg_m05} we display a series of histograms with the stellar population synthesis results (also summarized in Tab.~\ref{sr}) for the CNSFRs: Each panel represents the results for the respective region marked in green circles in our maps. From these plots it is clear the presence of young stars (7$-$10\,Myr) coexisting with an older SPC, which we attribute to the underlying bulge component (3/13\,Gyr). A small contribution of up to 20\% of intermediate-age stars with a large spread in age (0.3$-$2.0\,Gyr) is seen in the CNSFRs, with exception of apertures {\it E} and {\it F}. Our results point out to a more complex nuclear ring SFH, with multiple starburst episodes, in agreement with recent results \citep{ma18}.

Previous studies have found an age dating for the star clusters distributed along the circumnuclear ring-like structure in NGC\,4303. \citetalias{colina00} found the east UV knots (blue contours in our maps) are older ($\sim$10$-$25\,Myr) than the west knots ($\sim$5$-$7.5\,Myr), thus suggesting an age offset in the ring. 

Even though emission-line measurements done by \citetalias{riffelA16} show a trend suggesting an age sequence along the ring increasing from {\it A} to {\it H} \citepalias[see Figures\,4 and 9 from][]{riffelA16}, the distribution of CNSFRs and their age differences indicate a rather episodic star formation along the ring, being equally consistent with regions to the east being older than regions to the west, in agreement with \citetalias{colina00}. For example, from H$_2$/Br$\gamma$ ratios, they found regions {\it G} and {\it H} to be older than regions {\it A}$-${\it F}, in full agreement with our results, as can be seen in Fig.~\ref{reg_m05}. 

In Tab.~\ref{sr} column (12) we present the SFRs derived for all the apertures. We can see {\it Region A} displays the highest SFR (0.14\,M$_{\odot}$yr$^{-1}$) in the ring, co-spatial with high H$\alpha$ emission as presented by \citet{colina99}. The range of the SFRs in the circumnuclear ring goes from 0.002$-$0.14\,M$_{\odot}$yr$^{-1}$, displaying $\sim$1 order of magnitude higher than those presented in \citetalias[][]{riffelA16} using the Br$\gamma$ luminosity and assuming a constant rate. These discrepancies might be related to the fact that with the stellar population fitting we are probing different (older) starbursts than those which are effectively ionizing the gas. In fact, \citet{ma18} found evidence suggesting the SFH in nuclear rings are better described by models of multiple bursts of star-formation, rather than assuming a constant SFR. In this work, the authors derived the age of the circumnuclear ring of NGC\,4303 through SED fitting and found it to be around 3\,Gyr (using \citetalias{conroy09} and assuming Z=Z$_{\odot}$), in agreement with our results (see column (7) - Tab.~\ref{sr}).

In order to better compare our results with the literature, we divided the SFR by the area of the extractions, obtaining the star formation surface density ($\sum_{SFR}$, column (13) - Tab.~\ref{sr}). In the nucleus, $\sum_{SFR}$ reaches up to $\sim$40\,M$_\odot$yr$^{-1}$kpc$^{-2}$, while in the ring values are in the range of 0.5$-$13\,M$_\odot$yr$^{-1}$kpc$^{-2}$. Typical star formation surface density values range from 1$-$50\,M$_\odot$yr$^{-1}$kpc$^{-2}$ on hundreds of parsec scales, 50$-$500M$_\odot$yr$^{-1}$kpc$^{-2}$ on scales of tens of parsecs, while it increases up to $\sim$1000\,M$_\odot$yr$^{-1}$kpc$^{-2}$ on parsec scales \citep[][and references therein]{valencia12}. Our $\sum_{SFR}$ values for the CNSFRs, including the nucleus (inner $\sim$120\,pc) are within the first $\sum_{SFR}$ ranges. This is in agreement with previous results, such as those presented by \citet{busch17}, which found $\sum_{SFR}$ of $\sim$28\,M$_\odot$yr$^{-1}$kpc$^{-2}$ in the nucleus and 5$-$13\,M$_\odot$yr$^{-1}$kpc$^{-2}$ in the ring of the nearby barred galaxy NGC\,1808. Thus, confirming are results display typical star formation surface density values.

Based on the spatial distributions of CNSFRs in nuclear rings, for example, \citet{boker08} discussed two scenarios of star formation: the `pop-corn' scenario in which the stellar clusters form at random positions producing no systematic age gradients and, the `pearls-on-a-string' scenario in which the clusters are formed where the gas enters the ring and then age as they orbit the ring forming a string of aging clusters \citep[see also][]{ryder01,diaz07}. \citetalias{riffelA16} could not favor neither of the two scenarios from the emission-line study, suggesting stars formed quasi-simultaneously over a large sector of the ring, aging as they rotate with an orbital time of $\sim$10\,Myr. \citep{seo13} performed simulations on how star formation proceed in nuclear rings of barred galaxies and found a critical SFR (SFR$_{c}$) of $\sim$1\,M$_{\odot}$yr$^{-1}$ which determines how star formation takes place in these rings. When the SFR in the ring is low (<SFR$_{c}$), star formation mostly takes place in the contact points of the ring with dust lanes leading to a age gradient (`pearls-on-a-string'). Otherwise, if the SFR is high (>SFR$_{c}$), then star formation is randomly distributed along the ring (`pop-corn'). \citet{mazzuca08} found similar results studying 22 nuclear rings (with SFRs in the range of 01$-$10\,M$_{\odot}$yr$^{-1}$), but for a higher SFR$_{c}$ of $\sim$3\,M$_{\odot}$yr$^{-1}$. Summing up the SFRs of the CNSFRs (excluding the nuclear aperture) we ended up with a SFR$\sim$0.2\,M$_{\odot}$yr$^{-1}$, a typical value for nuclear rings \citep{ma18}. Even if we add the mean SFR of the northeast section of the ring ($\leq$2.5$\times$10$^{-3}$\,M$_{\odot}$yr$^{-1}$, see Fig.~\ref{maps_extra}), the SFR in the circumnuclear ring of NGC\,4303 would still be lower that the SFR$_{c}$ proposed by these previous results. Thus, our SFR results favor the `pearls-on-a-string' scenario, with a gradient of increasing age going from regions {\it A} to {\it H}.

When comparing the reddening values derived from the emission lines in \citetalias{riffelA16} with those found with our stellar population fitting, the former present higher A$_V$ (reaching up to 2.5\,mag in {\it Regions B} and {\it E}). This is expected and is related to the fact that the hot ionizing stars are associated to dustier regions with respect to the cold stellar population \citep{calzetti94}. The ring displays reddening values from 0$-$0.9\,mag, lower than typical values from previous studies ranging from 2$-$5\,mag \citep{krabbe94,kotilainen96,rosenberg12,busch17}, but consistent with UV and optical results from \citetalias[][]{colina00} for less obscured regions in the {\it V$-$H} colors, as well as with Cluster\,G of \citetalias[][]{colina02} with A$_V$=0.6\,mag (using R$_V$=4.05), which we call {\it Region A}. 

In addition, the Adev values in Fig.~\ref{reg_m05}, column (12) illustrate the good quality of the fits for the CNSFRs.

\subsection{Stellar populations in the nuclear region: Is there any evidence of the LLAGN?}\label{nucleus}

The presence of an AGN in the central region of NGC\,4303 is still a matter of debate. Using NIR emission-line ratios, \citetalias{riffelA16} constructed a NIR diagnostic diagram \citep[proposed by][]{colina15}, which suggested the presence of an AGN at the nucleus, confirming previous results that the nuclear emission of this galaxy has a composite nature \citep[LLAGN plus a young massive stellar cluster - e.g., ][]{colina02,bailon03}. 

As already mentioned in Sec.~\ref{intro}, previous studies suggested a young (4\,Myr) massive SSC dominates the UV emission, while an intermediate/old (1$-$5Gyr) stellar population dominates the optical continuum \citepalias{colina00,colina02}. Our results in the nuclear region (inner 60\,pc radius) of this source show the same trend, presenting a contribution of young stars with rather older ages (7$-$8.5\,Myr) and a rather younger intermediate-age SPC (0.3$-$0.7\,Gyr) plus a dominating ($\sim$50\%) 13\,Gyr old stellar population, which we attribute to the underlying bulge. Moreover, \citetalias{colina00} suggested the {\it H} excess in the inner 8\,pc could be related to the presence of luminous red supergiants, implying a second star formation episode with $\sim$10\,Myr, which are close in age with the young stars we found. These authors also argued the large {\it V$-$H} colors (3.2\,mag) found in this inner region could not be fully accounted for by small amounts of dust extinction and suggested the presence of an extremely red source, such as a hidden AGN. Assuming a pure power law with $\nu$=-1, representing the AGN featureless emission, they obtained values of {V$-$H}=2.3\,mag, closer to the observed ones, but still not fully compatible. 

From the stellar population synthesis we do not find contribution of the featureless continuum and/or hot emission component to the inner 60\,pc radius. Although this result does not rule out the existence of a hidden AGN in the centre of NGC\,4303, it supports the scenario of a LLAGN/LINER-like source rather than a Seyfert\,2 nucleus \citep{colina99}. In a recent work, \citet{burtscher15} found no evidence of nuclear EW dilution of the CO feature by AGN light in this source (as can be seen in their figure\,2.), in agreement with our stellar population analysis. It is important to call attention to the fact that the spectra of very young SSPs \citep[$t\leq 5\,Myr$,][ and references therein]{riffel09} and the featureless continuum emission can be degenerate in the fits, meaning we cannot fully discriminate between these two components using our methodology. Nevertheless, we are favorable to believe the young SPC we are finding with stellar population synthesis is real, since there is more than sufficient evidence in the literature of the presence of young stars in the nuclear region of this source. Furthermore, our fits find SSPs in the range of 7$-$8.5\,Myr, which are not the younger SSPs ($t \lesssim$5\,Myr) which usually mimic the contribution of a featureless continuum \citep{cid04}.

Additional support for the scenario of a LLAGN/LINER-like source rather than a Seyfert\,2 nucleus comes from the fact that the inclusion of a featureless continuum (power law) and hot dust emission (blackbody functions) has been used to trace signatures of luminous AGNs by our AGNIFS group when performing stellar population synthesis in the NIR, and in the case of Seyfert sources these components are required to properly fit their spectral energy distribution.


\subsubsection{Nuclear extinction}

Typical nuclear reddening values (A$_V$) for Seyfert galaxies derived using NIR emission-lines are in the range of 1.3$-$5\,mag for Type\,1 nuclei and 1.8$-$9.0\,mag for Type\,2 \citep[][and references therein]{valencia12}. These authors also derived a 2.5\,mag reddening for the nuclear region of a starburst/Seyfert composite galaxy. In the same line, \citet{dametto14} found nuclear A$_V$ ranging from 2.5$-$8.0\,mag for the gas in four Starburst galaxies, while for the stars they found 1.0$-$3.2\,mag. In addition, AGNIFS group derived nuclear reddening values using NIR data from Seyfert\,2 galaxies and found values between 2$-$4\,mag for the stars, while for the gas these values reach up to 7.0\,mag in Mrk\,573 \citep{riffelA10,riffel11c,diniz17}.     

Previous reddening estimations were made for NGC\,4303 in \citetalias{colina02} using the emission lines as well as the UV shape of the spectrum for the nuclear region. These authors found A$_V$=0.2/0.4\,mag using UV spectra (LMC/Calzetti extinction law), while using H$\alpha$/H$\beta$ line ratio they fitted a dust free nuclear spectrum (r=0\farcs45$\sim$35\,pc) for this object. These low extinction values are in agreement with our results for the nuclear extraction (r=0\farcs75$\sim$60\,pc), as can be seen in the top panel of Fig.~\ref{reg_m05}. However, when looking to the results spaxel-by-spaxel in Fig.~\ref{maps_extra} (left panel), we derived rather higher $A_V$ values ($\sim$2.3\,mag) for the inner tens of parsecs in this source, which is within the typical reddening values. \citetalias{colina00} had found large {\it V$-$H} colors (3.2\,mag) for the inner regions (<8\,pc) of NGC\,4303, as already mentioned, arguing this could possibly be related to a hidden AGN, which is consistent with our results (see Sec.~\ref{nucleus}). 

\subsubsection{Nuclear SFRs}

In Tab.~\ref{sr} we present SFR values for the nuclear region (0.43\,M$_{\odot}$yr$^{-1}$), which are in the range of typical values found in the literature (see also Sec.~\ref{cnsfrs}). For example, using Br$\gamma$ luminosity, \citet{valencia12} and \citet{busch17} derived nuclear SFRs of 0.18\,M$_{\odot}$yr$^{-1}$ and $<$ 0.35\,M$_{\odot}$yr$^{-1}$, respectively, in the inner 57/50\,pc radius in their sample. \citet{bailon03} and \citetalias{riffelA16} also calculated the SFR in the inner hundreds of parsecs in NGC\,4303 using emission line indicators in the optical and NIR, respectively, and both found 0.013\,M$_{\odot}$yr$^{-1}$, a lower value than the one we are finding. As discussed in Sec.~\ref{cnsfrs}, this might be related to the different ages of the bursts we are probing with stellar populations synthesis, which considers a wide range of bursts up to 10\,Myr. When using emission line indicators to derive SFRs, one is looking into a more instantaneous burst scenario, favoring the youngest components of the stellar population.

Assuming the SFR of 0.43\,M$_{\odot}$yr$^{-1}$ for the last 10\,Myr, the inner 60\,pc radius of this source would have formed $\sim$4$\times$10$^{6}$M$_{\odot}$. Then, if we assume an efficiency of $\sim$0.1 to convert gas into stars, this region should comprise around 10$^7$\,M$_{\odot}$ of cold molecular (H$_2$) gas (M$_{cold}$), which is in agreement with the values found by previous studies for the cold molecular gas in this region of NGC\,4303, such as \citet{schinnerer02} and \citetalias{riffelA16}. The former derived M$_{cold}$=6.9$\times$10$^7$M$_{\odot}$ directly from CO observations of the nuclear disc of this object, while the later estimated M$_{cold}$=6.25$\times$10$^6$M$_{\odot}$ from converting hot to cold molecular gas masses \citep[e.g.][]{mazzalay13}.

A alternative way to approach this would be to compare rates of star formation, mass-inflow and mass accretion to the black hole. Mass-inflow rates are in the range of 10$^{-2}-$10\,M$_{\odot}$yr$^{-1}$, which is consistent with our nuclear SFR (0.43\,M$_{\odot}$yr$^{-1}$) plus a typical mass accretion rate to the black hole of 10$^{-3}$ to 10$^{-2}$\,M$_{\odot}$yr$^{-1}$ \citep[][and references therein]{riffelA13b}. Thus, the values we are finding for the SFR in the central region of this source are feasible with the amount of gas available to form stars. 


\section{conclusions}\label{conclusions}

We present the first spatially resolved stellar population study of the inner $\sim$200\,pc radius of NGC\,4303 in the NIR. Using J-, H- and K-band SINFONI/VLT data, stellar population synthesis was performed with the {\sc starlight} code and \citetalias{maraston05} SSP models. The main conclusions of this work go as follows:

(i) The dominant stellar population component presents a spatial variation in the inner $\sim$200\,pc radius of this source, suggesting an age stratification. The youngest stellar population components (t $\leq$ 2\,Gyr) are distributed along a circumnuclear ring with 200$-$250\,pc radius in agreement with previous studies \citepalias{riffelA16}. Three main components stand out: two nuclear blobs, one composed by young stars (t $\leq$ 50\,Myr) and shifted towards east from the centre (here defined as the peak of the Br$\gamma$ emission line) and one composed by intermediate-age stars (50\,Myr $<$ t $\leq$ 2\,Gyr) located southwest from the centre; and an internal spiral arm-like structure also composed by intermediate-age stars surrounding the blob of young stars. The old stellar population component which we attribute to an underlying bulge stellar population is distributed outside the two blob structures, with an enhanced contribution northwest from the centre. These results reveal a rather complex star formation history in NGC\,4303, indicating star formation has occurred through multiple bursts in this source.  

(ii) With our stellar population synthesis analysis we favor the interpretation of a circumnuclear ring of star formation in the inner $\sim$250\,pc radius of NGC\,4303 rather than a spiral arm structure as suggested by \citetalias{colina02}, in agreement with the analysis of the emission gas presented in \citetalias{riffelA16}. We suggest the spiral arm seen in UV images is tracing the less obscured star-forming regions of the circumnuclear ring.

(iii) Circumnuclear star-forming regions (CNSFRs) distributed along the ring present SFRs in the range of 0.002$-$0.14\,M$_{\odot}$yr$^{-1}$. Our results favor the `pearls-on-a-string' star formation scenario with an age gradient along the ring. This conclusion is based on the fact that we have found a value for the total SFR in the ring which is lower than the critical SFR ($\sim$1\,M$_{\odot}$yr$^{-1}$) to separate between the `pearls-on-a-string' and the `pop-corn' scenarios.

(iv) At the nuclear region (R$\lesssim$ 60\,pc) we find a series of star formation bursts: a first one occurring 13\,Gyrs ago ($x_o$), accounting for 50\% of the light at 2.067$\mu$m; a set of individual minor bursts with ages from 0.3$-$0.7\,Gyr ($x_i$) corresponding to $\sim$15\% of the light and a contribution of $\sim$35\% of the $x_y$ component with a major burst ($\sim$ 30\%) at 7.5\,Myr. We derive a nuclear SFR of 0.43\,M$_{\odot}$yr$^{-1}$, corresponding to a star formation surface density ($\sum_{SFR}$) of $\sim$40\,M$_\odot$yr$^{-1}$kpc$^{-2}$.

(v) No signatures of non-thermal featureless continuum and hot dust (blackbody functions) components were necessary to reproduce the nuclear (inner 60\,pc) continuum of this source. This result supports the scenario in which a LLAGN/LINER-like source is hidden in the centre of NGC\,4303, rather than a Seyfert\,2 nucleus. In the case of Seyferts, such components are required to properly reproduce their spectral energy distribution. 

(vi) A reddening free nuclear spectrum (R$\lesssim$ 60\,pc) was fitted for this source, in agreement with UV and optical previous results. When looking to the most inner (unresolved) regions ($<$10\,pc), we find a reddened solution consistent with the presence of a hidden AGN, as already proposed by \citetalias{colina00} from the large {\it V$-$H} colors (3.2\,mag) in this region.  

(vii) From our spatially resolved comparison of the results obtained using different SSP models, we concluded the use of \citetalias{maraston05} models, in general, yields a more consistent scenario for the distribution of the stellar populations. Results using \citetalias{bc03} and \citetalias{conroy09} models miss the contribution of intermediate-age stars, which were expected to be found in active star formation sources. We speculate the main reason for discrepancies found between the results using these three low-resolution EPS models sets is related to their prescription for implementing the TP-AGB phase to the models.  

As LLAGNs are the most common type of nuclear activity in the Local Universe, the understanding of their nature and SFH is of utmost importance for the comprehension of galaxy formation and evolution. Our results on the spatially resolved stellar populations of the few hundreds of parsecs in the nearby LLAGN NGC\,4303 show a complex star formation scenario, with different stellar population components coexisting with a low efficiency accreting black hole. Thus, this detailed study help to shed some light on the contemporary scenarios of galaxy formation and evolution.

\section*{Acknowledgements}


We thank an anonymous referee for useful suggestions which helped to improve the paper.
The authors would like to thank Dr. Aberto Rodr\'iguez Ardila of the Laborat\'orio Nacional de Astronomia, Brazil, and Dr. Lucimara Martins of Universidade Cruzeiro do Sul, Brazil, for kindly sharing with us their IRTF SpeX spectrum of NGC\,4303. 
The Brazilian authors acknowledge support from FAPERGS (Funda\c{c}\~ao de Amp\'aro \`a pesquisa do Estado do Rio Grande do Sul) and CNPq (Conselho Nacional de Desenvolvimento Cient\'ifico e Tecnol\'ogico).
L.C. acknowledges support from CNPq special visitor fellowship PVE 313945/2013-6 under the Brazilian program Science without Borders. 
L.C. and S.A. are supported by grants AYA2012-32295, AYA2012- 39408,  ESP2015-68964-P, and ESP2017-83197-P, and J.P. is supported by grant AYA2017-85170-R, all from the Ministerio de Econom\'ia y Competitividad of Spain.
The {\sc starlight} project is supported by the Brazilian agencies CNPq, CAPES and FAPESP (Funda\c{c}\~ao de Amp\'aro \`a pesquisa do Estado de S\~ao Paulo) and by the France-Brazil CAPES/Cofecub program.




\bibliographystyle{mnras}
\bibliography{refs} 



%
%
%
%
%

\bsp        
\label{lastpage}
\end{document}